\documentclass[twocolumn,superscriptaddress,aps,pre]{revtex4-2}

\usepackage[utf8]{inputenc}
\usepackage{url}
\usepackage{amssymb,amsmath,amsthm}
\usepackage{bm}
\usepackage{graphicx}
\usepackage{enumerate}
\usepackage{microtype}
\usepackage{float}
\usepackage{color}
\usepackage{hyperref}
\usepackage{multirow}
\usepackage{array}
\newcolumntype{C}[1]{>{\centering\let\newline\\\arraybackslash\hspace{0pt}}m{#1}}

\usepackage[ruled,linesnumbered,vlined]{algorithm2e}

\begin{document}

\title{Explosive rigidity percolation in origami}

\author{Rongxuan Li}
\affiliation{Department of Mathematics, The Chinese University of Hong Kong}
\author{Gary P. T. Choi}
\thanks{To whom correspondence may be addressed. Email: ptchoi@cuhk.edu.hk.}
\affiliation{Department of Mathematics, The Chinese University of Hong Kong}
\date{\today}

\begin{abstract}
Origami, the traditional art of paper folding, has revolutionized science and technology in recent years and has been found useful in various real-world applications. In particular, origami-inspired structures have been utilized for robotics and mechanical information storage, in both of which the rigidity control of origami plays a crucial role. However, most prior works have only considered the origami design problem using purely deterministic or stochastic approaches. In this paper, we study the rigidity control of origami using the idea of explosive percolation. Specifically, to turn a maximally floppy origami structure into a maximally rigid origami structure, one can combine a random sampling process of origami facets and some simple selection rules, which allow us to exploit the power of choices and significantly accelerate or delay the rigidity percolation transition. We further derive simple formulas that connect the rigidity percolation transition effects with the origami pattern size and the number of choices, thereby providing an effective way to determine the optimal number of choices for achieving prescribed rigidity percolation transition accuracy and sharpness. Altogether, our work paves a new way for the rigidity control of mechanical metamaterials.

\end{abstract}

\maketitle

\section{Introduction}
Origami, the traditional art of paper folding, has existed in various regions and cultures for centuries~\cite{hatori2011history}, in which most paper folding practices were primarily related to religious, ceremonial, and recreational purposes. In the past several decades, origami has been gaining popularity among not just artists but also scientists and engineers for its rich geometrical and mechanical properties. In particular, the mathematics and computation of origami have been extensively studied by different origami theorists~\cite{huffman1976curvature,miura1985method,hull1994mathematics,lang1996computational,kawasaki2005roses,demaine2007geometric,lang2012origami}. Moreover, origami has been utilized in many modern engineering and technological applications, including flexible electronics~\cite{nogi2013foldable,zhang2017origami}, soft robotics~\cite{rafsanjani2018kirigami,ze2022soft} and space exploration~\cite{nishiyama2012miura}. Several surveys on recent advances in origami and its applications can be found in~\cite{bertoldi2017flexible,surjadi2019mechanical,meloni2021engineering,rus2018design,zhai2021mechanical,choi2024computational,misseroni2024origami}.

Many prior origami research works have focused on the geometrical and physical properties of various classical origami patterns, such as the Miura-ori pattern~\cite{wei2013geometric,schenk2013geometry}, waterbomb origami~\cite{hanna2014waterbomb}, Resch pattern~\cite{tachi2013designing} and curved crease origami~\cite{dias2012geometric}. Also, multiple computational algorithms have been developed to modify the geometry of origami structures to achieve different desired properties~\cite{tachi2009origamizing,dudte2016programming,dudte2021additive,dang2022inverse}. In recent years, there has been an increasing interest in the mechanical properties of different origami structures~\cite{liu2017nonlinear,chen2018branches,he2019rigid,he2020rigid,he2022rigid,zhang2023rigidity} and the application of origami in the design of mechanical storage devices~\cite{treml2018origami,meng2021bistability}. In~\cite{chen2019rigidity}, Chen and Mahadevan studied the rigidity percolation and geometric information in floppy origami. Specifically, they changed the planarity of the facets in the Miura-ori structure using a stochastic approach and analyzed the resulting changes in its rigidity.

\begin{figure}[t]
    \centering
    \includegraphics[width=\linewidth]{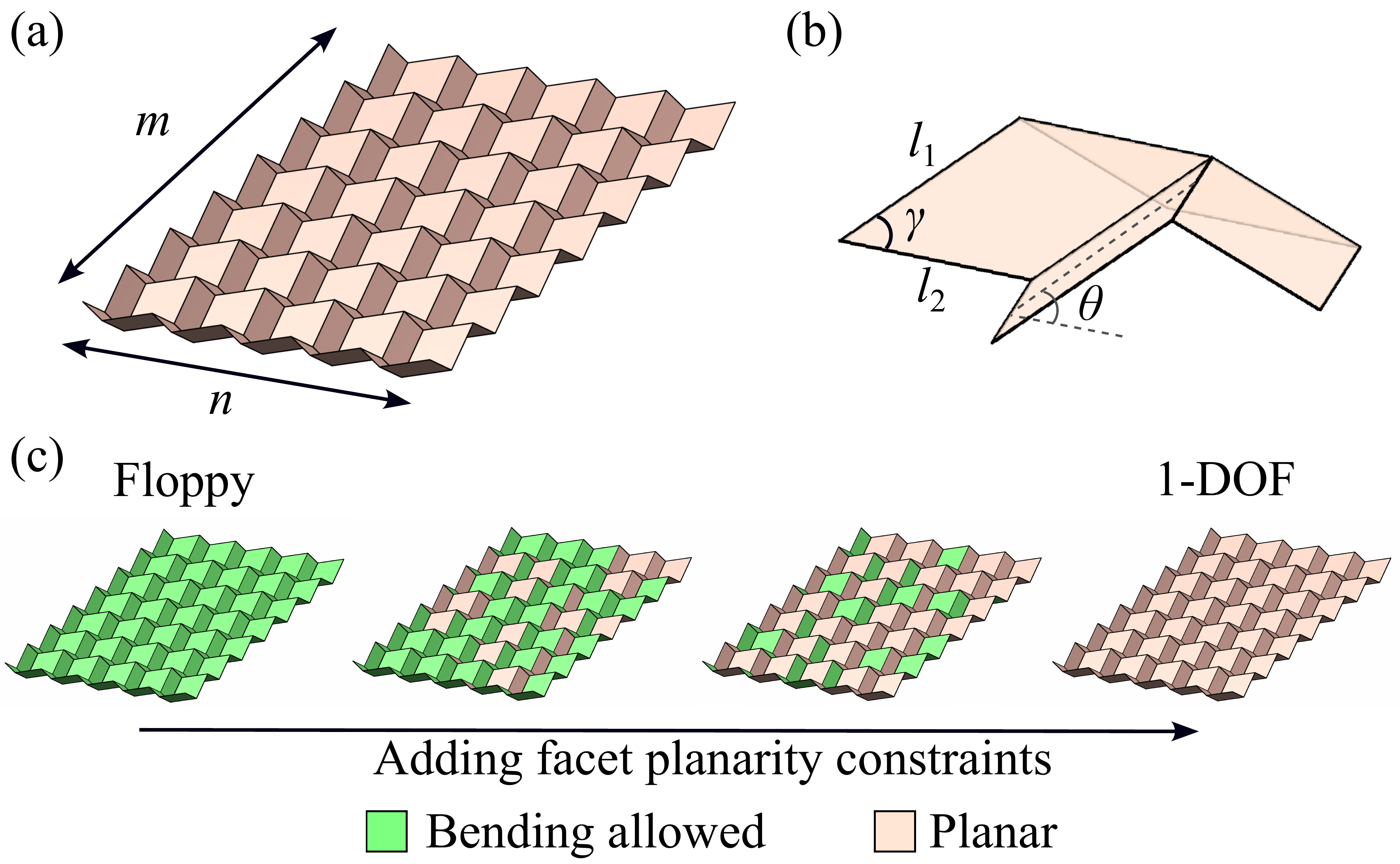}
\caption{\textbf{The Miura-ori pattern and rigidity percolation.} (a)~A Miura-ori structure with $m$ rows of facets in the vertical direction and $n$ columns of facets in the horizontal direction. (b)~A Miura-ori unit cell with four facets is characterized by two angles $\gamma$ and $\theta$ and two length parameters $l_1$ and $l_2$. (c)~Starting from a maximally floppy origami structure in which all facets are allowed to bend out-of-plane (left), one can gradually impose planarity constraints on certain facets, thereby changing the rigidity of the overall origami structure. In particular, note that explicitly imposing the planarity constraint on a facet may affect not just itself but also some other facets implicitly. As all facets become planar, the resulting origami structure has exactly 1 degree of freedom (DOF) associated with a rigid folding motion.}
    \label{fig:illustration}
\end{figure}

Note that the rigidity percolation in bond networks has been widely studied over the past few decades~\cite{jacobs1996generic,ellenbroek2011rigidity,zhang2015rigidity}. In recent years, explosive percolation in random networks has received a lot of attention~\cite{achlioptas2009explosive,radicchi2009explosive,ziff2010scaling,araujo2010explosive,da2010explosive}. Specifically, it has been shown that incorporating an extra step of selecting between two choices in the random process (also known as \emph{the power of two choices}) can lead to a very sharp phase transition. More recently, the idea of explosive percolation has been applied to the rigidity and connectivity control in kirigami metamaterials~\cite{chen2020deterministic,choi2023explosive}. Motivated by the above works, here we pose and solve the problem of optimally controlling the rigidity percolation transition in origami using simple selection rules based on the idea of explosive percolation.

\section{Methods}

The Miura-ori pattern is an array of quadrilaterals consisting of $m$ rows of facets in the vertical direction and $n$ columns of facets in the horizontal direction (Fig.~\ref{fig:illustration}(a)). It is composed of identical unit cells with four facets, which can be characterized by two angle parameters, $\gamma$ and $\theta$, and two length parameters, $l_1$ and $l_2$~\cite{wei2013geometric,schenk2013geometry} (Fig.~\ref{fig:illustration}(b)). For the classical Miura-ori pattern, all quadrilateral facets are planar, and the structure has only one degree of freedom (DOF), excluding the three global translational DOFs and three global rotational DOFs. Specifically, it features a single zero-energy DOF associated with a rigid folding motion~\cite{miura1985method}.

Note that if we allow every quadrilateral facet to bend along one of its diagonals, the origami structure becomes floppy (Fig.~\ref{fig:illustration}(c)). It is natural to ask how the rigidity of the structure evolves as we start from the initial maximally floppy state and gradually enforce that the facets remain planar, thereby preventing out-of-plane folding. In particular, it is noteworthy that explicitly enforcing a facet to be planar may affect not just its planarity but also implicitly the planarity of some other facets. Therefore, the rule of adding the planarity constraints throughout the process has a significant effect on the rigidity change. 

We first consider imposing certain geometrical constraints as described in~\cite{chen2019rigidity} and constructing the infinitesimal rigidity matrix of the origami structure to determine the range of motions associated with infinitesimal rigidity. More specifically, we consider the following \emph{edge constraint} for each edge $(\mathbf{v}_i, \mathbf{v}_j)$ to enforce that all edge lengths remain unchanged:
\begin{equation} \label{eqt:edge_length_constraint}
    g_e = \|\mathbf{v}_i - \mathbf{v}_j\|^2 - l_e^2 = 0,
\end{equation}
where $l_e = l_1$ or $l_2$ is the edge length of the quadrilateral facets. Note that there are in total $m(n+1)+(m+1)n = 2mn+m+n$ edges in a $m\times n$ Miura-ori structure, and hence we have $2mn+m+n$ edge constraints (which can be denoted as $g_{e_1}, g_{e_2}, \dots, g_{e_{2mn+m+n}}$).

To prevent shearing of the quads, for each quad $(\mathbf{v}_{i_1},\mathbf{v}_{i_2},\mathbf{v}_{i_3},\mathbf{v}_{i_4})$ (where the four vertices represent the bottom left, bottom right, top right and top left vertices respectively by our convention), we have the following \emph{diagonal (no-shear) constraint} for one of the diagonals (for consistency we always choose the diagonal $(\mathbf{v}_{i_1},\mathbf{v}_{i_3})$ involving the bottom left vertex $\mathbf{v}_{i_1}$ and the top right vertex $\mathbf{v}_{i_3}$):
\begin{equation} \label{eqt:diagonal_constraint}
g_d = \|\mathbf{v}_{i_3} - \mathbf{v}_{i_1}\|^2 - l_d^2 = 0,
\end{equation}
where $l_d$ is the diagonal length of the quadrilateral facets. Note that there are in total $mn$ facets in a $m\times n$ Miura-ori structure, and hence we have $mn$ diagonal constraints (which can be denoted as $g_{d_1}, g_{d_2}, \dots, g_{d_{mn}}$).

We will also add the following \emph{quad planarity constraint} gradually to enforce the planarity of some specific quadrilateral facets. Specifically, consider a quad $(\mathbf{v}_{i_1},\mathbf{v}_{i_2},\mathbf{v}_{i_3},\mathbf{v}_{i_4})$. Suppose there is a virtual diagonal edge $(\mathbf{v}_{i_1},\mathbf{v}_{i_3})$ in the quad that allows the quad to fold about it. Then, to enforce the quad to be planar, the volume of the tetrahedron formed by $\mathbf{v}_{i_1},\mathbf{v}_{i_2},\mathbf{v}_{i_3},\mathbf{v}_{i_4}$ must be 0. Therefore, we have:
\begin{equation} \label{eqt:quad_planarity_constraint}
    g_p = (\mathbf{v}_{i_2} - \mathbf{v}_{i_1}) \times (\mathbf{v}_{i_4} - \mathbf{v}_{i_1}) \cdot (\mathbf{v}_{i_3} - \mathbf{v}_{i_1}) = 0.
\end{equation}

We can then construct the infinitesimal rigidity matrix as described in~\cite{guest2006stiffness} to examine the possible infinitesimal modes of motion. Specifically, suppose there are in total $2mn+m+n$ edge constraints, $mn$ diagonal constraints, and $t$ planarity constraints imposed. Denote $K = (2mn+m+n) + mn + t$ as the total number of constraints and $V = (m+1)(n+1)$ as the total number of vertices in the $m\times n$ Miura-ori structure. The infinitesimal rigidity matrix is a $K \times V$ matrix $A$ with 
\begin{equation}
    A = \begin{pmatrix}
        \frac{\partial g_1}{\partial x_1} &  \frac{\partial g_1}{\partial y_1}  &  \frac{\partial g_1}{\partial z_1} & \frac{\partial g_1}{\partial x_2} &  \frac{\partial g_1}{\partial y_2}  &  \frac{\partial g_1}{\partial z_2} & \dots & \frac{\partial g_1}{\partial z_{V}}\\
        \frac{\partial g_2}{\partial x_1} &  \frac{\partial g_2}{\partial y_1}  &  \frac{\partial g_2}{\partial z_1} & \frac{\partial g_2}{\partial x_2} &  \frac{\partial g_2}{\partial y_2}  &  \frac{\partial g_2}{\partial z_2} & \dots& \frac{\partial g_2}{\partial z_{V}}\\
        \vdots & \vdots & \vdots & \vdots & \vdots & \vdots & \ddots & \vdots\\
        \frac{\partial g_K}{\partial x_1} &  \frac{\partial g_K}{\partial y_1}  &  \frac{\partial g_K}{\partial z_1} & \frac{\partial g_K}{\partial x_2} &  \frac{\partial g_K}{\partial y_2}  &  \frac{\partial g_K}{\partial z_2} & \dots& \frac{\partial g_K}{\partial z_{V}}
    \end{pmatrix},
\end{equation}
where $g_1, g_2, \dots, g_K$ include all edge constraints $\{g_{e_i}\}_{i=1}^{2mn+m+n}$, all diagonal constraints $\{g_{d_i}\}_{i=1}^{mn}$, and the current set of quad planarity constraints $\{g_{p_i}\}_{i=1}^{t}$, and $(x_i, y_i, z_i)$ are the coordinates of the vertex $\mathbf{v}_i$, $i = 1, 2, \cdots, V$.

It is easy to see that the matrix $A$ is a sparse matrix and all entries of it can be explicitly derived. For instance, for each edge constraint in Eq.~\eqref{eqt:edge_length_constraint}, suppose $\mathbf{v}_i = (x_i, y_i, z_i)$ and $\mathbf{v}_j = (x_j, y_j, z_j)$. The partial derivatives of $g_e$ with respect to the coordinate variables are given by:
\begin{align}
    \frac{\partial g_e}{\partial x_i} &= -\frac{\partial g_e}{\partial x_j} = 2(x_i-x_j),\\
    \frac{\partial g_e}{\partial y_i} &= -\frac{\partial g_e}{\partial y_j} = 2(y_i-y_j),\\
    \frac{\partial g_e}{\partial z_i} &= -\frac{\partial g_e}{\partial z_j} = 2(z_i-z_j),
\end{align}
and the partial derivatives of $g_e$ with respect to all other variables are 0. This shows that each row of $A$ associated with an edge constraint has at most 6 non-zero entries. See Appendix~\ref{sec:appendix_matrix} for the explicit formulas for all other entries of $A$.

Now, suppose there is an infinitesimal displacement $\overrightarrow{dv}$ added to all vertex coordinates $\vec{v} = [x_1, y_1, z_1, x_2, y_2, z_2, \dots, x_V, y_V, z_V]^T$. The condition for infinitesimal rigidity is given by
\begin{equation}
    A \overrightarrow{dv} = 0.
\end{equation}
Therefore, the infinitesimal DOF of the origami structure is the dimension of the null space of $A$. In other words, we can calculate the infinitesimal DOF (with the three global translational DOFs and the three global rotational DOFs removed) by:
\begin{equation}\label{eqt:DOF}
    d = 3(m+1)(n+1) - \text{rank}(A) - 6.
\end{equation}

\begin{figure*}[t!]
    \centering
    \includegraphics[width=\linewidth]{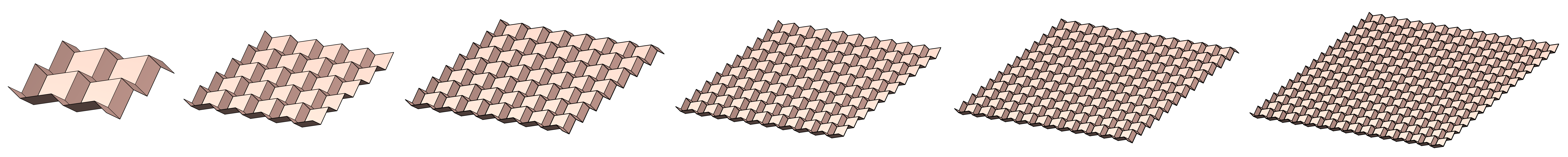}
    \caption{\textbf{Miura-ori structures of different sizes.} Left to right: $L\times L = 5\times 5, 10 \times 10, 15 \times 15, 20 \times 20, 25 \times 25, 30 \times 30$. The structures are not displayed to scale.}
    \label{fig:size}
\end{figure*}

As described in~\cite{chen2019rigidity}, the infinitesimal rigidity matrix $A$ for the initial maximally floppy Miura-ori structure only involves the $2mn+m+n$ edge constraints and the $mn$ diagonal constraints. Therefore, the DOF of the initial structure is $d_{\text{initial}} = 3(m+1)(n+1) - (2mn+m+n+mn) - 6 = 2m+2n - 3$, while the DOF of the final Miura-ori structure after all facets are forced to be planar is 1. We can then study the change of DOF from $2m+2n-3$ to 1 as the quad planarity constraints are gradually imposed, where at each step we choose a new quad and explicitly impose the quad planarity constraint on it. Specifically, we define the planarity constraint density $\rho \in [0, 1]$ as follows:
\begin{equation}
    \rho = \frac{\text{\#(quad planarity constraints imposed)}}{mn}.
\end{equation}
Since adding each quad planarity constraint will increase the number of rows of the matrix $A$ by 1, it follows that the DOF $d$ in Eq.~\eqref{eqt:DOF} will either be unchanged or reduced by 1. Therefore, to get a 1-DOF structure, the minimum number of quad planarity constraints needed is $(2m+2n-3)-1 = 2m+2n-4$. In other words, the theoretical minimum density $\rho_{\min}$ for getting a 1-DOF structure is 
\begin{equation}\label{eqt:min_theoretical}
    \rho_{\min} = \frac{2m+2n-4}{mn}.
\end{equation}
Also, note that the origami structure will not be 1-DOF unless all corner facets are planar. By deferring the addition of the quad planarity constraints on them to the end of the process, one can avoid getting a 1-DOF structure as much as possible. Therefore, the theoretical maximum density $\rho_{\max}$ for getting a 1-DOF structure is 
\begin{equation}\label{eqt:max_theoretical}
    \rho_{\max} = 1.
\end{equation}

Now, following the idea of explosive percolation~\cite{achlioptas2009explosive,radicchi2009explosive}, we consider gradually adding the quad planarity constraint to the initial floppy origami structure based on some selection rules to control the rigidity percolation transition. Let $k \geq 1$ be a positive integer. At each step, we sample $k$ facets randomly from the set of available facets. We then choose one among them based on one of the following selection rules:
\begin{itemize}
    \item \emph{Most Efficient selection rule}: Denote the $k$ randomly sampled candidate facets as $f_1, f_2, \dots, f_k$. For each facet $f_i$, we construct an augmented infinitesimal rigidity matrix $A_i$ by adding the quad planarity constraint imposed on the facet $f_i$ to the current infinitesimal rigidity matrix $A$. We then compute the DOF $d_i$ of the temporarily updated origami structure using Eq.~\eqref{eqt:DOF}. Among all $k$ candidate facets, we choose the facet that gives the minimum DOF, i.e. the facet $f_c$ with $c = \text{argmin}_i d_i$. If there are multiple facets that give the minimum DOF, we choose one among them randomly.
    \item \emph{Least Efficient selection rule}: Analogous to the above rule, we construct an augmented infinitesimal rigidity matrix $A_i$ for each candidate facet $f_i$ and compute the DOF $d_i$ of the temporarily updated origami structure. We then choose the facet that gives the maximum DOF among all $k$ choices, i.e. the facet $f_c$ with $c = \text{argmax}_i d_i$. If there are multiple facets that give the maximum DOF, we choose one among them randomly.
\end{itemize}
After choosing a facet among the $k$ candidate facets, we impose the quad planarity constraint on it and update the infinitesimal rigidity matrix $A$. We then repeat the above process until all the quad planarity constraints are explicitly imposed on all $mn$ facets. Here, note that after the $(mn-k)$-th step, there will be less than $k$ available facets. In this case, all available facets will be automatically considered. Also, our subsequent experiments and analyses involve simulations with various pattern sizes $mn$ and number of choices $k$ from a fixed list of values. By our convention, if the prescribed $k$ is larger than $mn$, we automatically correct it as $k = mn$.

\begin{figure*}[t!]
    \centering
    \includegraphics[width=\linewidth]{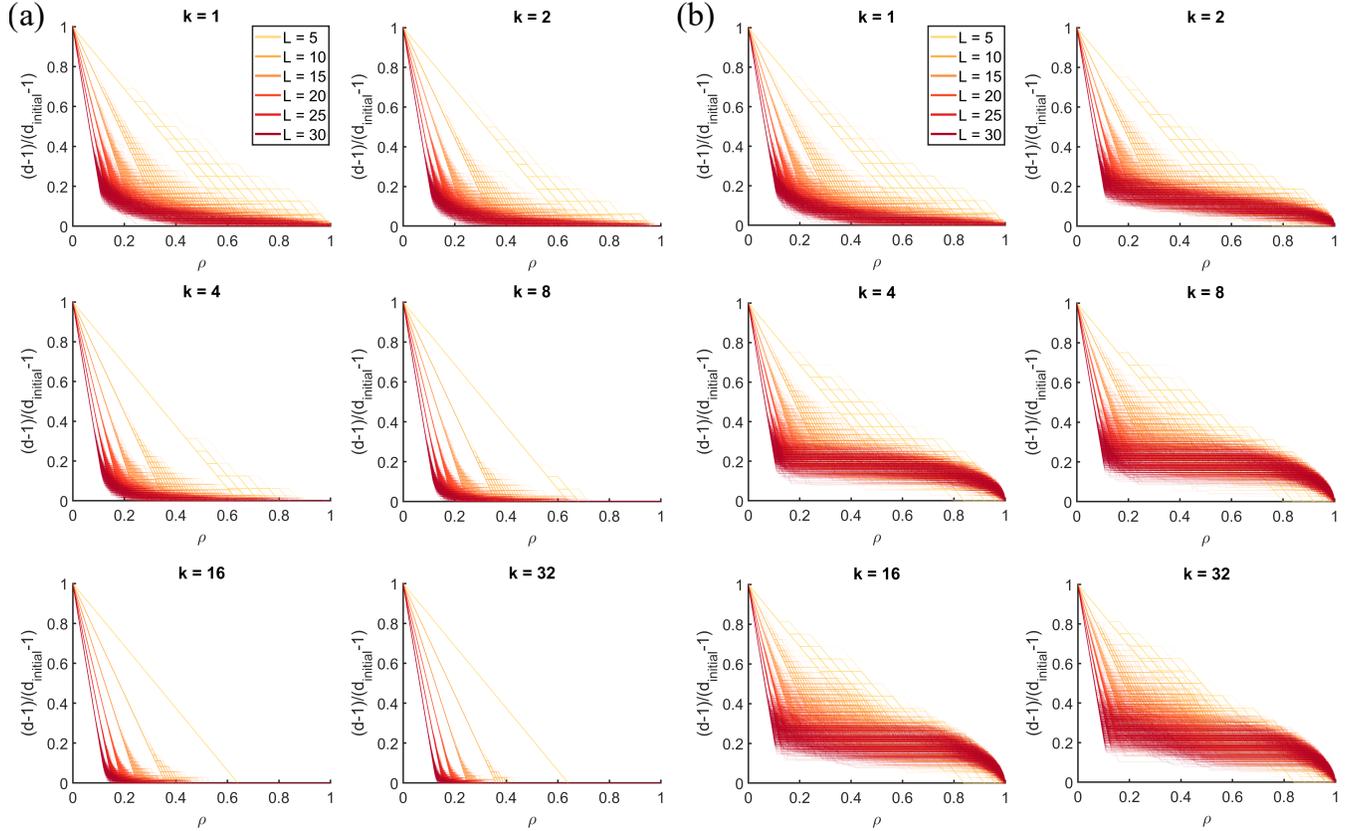}
\caption{\textbf{The change in the normalized DOF under different selection rules with different number of choices for $L\times L$ Miura-ori structures.} (a) The Most Efficient selection rule. (b) The Least Efficient selection rule. For each $k = 1, 2, 4, 8, 16, 32$, we plot the normalized DOF $\widetilde{d} = (d-1)/(d_{\text{initial}}-1)$ in all 500 simulations for all $L = 5, 10, 15, 20, 25, 30$ on the same plot to visualize the change in $\widetilde{d}$. Each partially transparent curve represents one simulation, and the opacity is proportional to the number of repeated trends. }
    \label{fig:result_square_dof_k}
\end{figure*}

\section{Results}
With the above formulation, it is natural to ask how the selection rules and the number of choices $k$ affect the rigidity percolation transition. To answer this question, we performed numerical simulations with different setups and analyzed the results. The numerical simulations were performed in MATLAB, with the Parallel Computing Toolbox used to improve the simulation efficiency. The infinitesimal rigidity matrix $A$ was constructed using the sparse matrix format in MATLAB. For the DOF calculation, we followed the approach in~\cite{chen2020deterministic} and used the built-in column approximate minimum degree permutation \texttt{colamd} and the QR decomposition \texttt{qr} functions to obtain the QR decomposition of $A$, and then approximated the rank of $A$ by counting the non-zero diagonal entries of the triangular matrix $R$. As the rank approximation of large matrices may be affected by numerical errors, we further restrict the DOF values to be within the feasible range $[1, d_{\text{initial}}]$. As for the geometry of the Miura-ori structure, we followed~\cite{chen2019rigidity} and used the length parameters $l_1 = l_2 = 2$ and angle parameters $\gamma = \pi/4$ and $\theta = \cos^{-1} \sqrt{2/3}$ to construct the Miura-ori unit cell. In Appendix~\ref{sec:appendix_geometry}, we present additional experiments to compare the simulations based on different geometric parameter setups, and the results show that the explosive percolation transition is independent of the geometry of the Miura-ori structure. 

\subsection{Explosive rigidity percolation in $L\times L$ Miura-ori}
To simplify our analysis, we first consider the case $m=n =L$ where $L$ is the number of rows/columns of quads in the Miura-ori structure. For each pattern size $L =  5, 10, 15, 20, 25, 30$ (see Fig.~\ref{fig:size}), each number of choices $k = 1, 2, 4, 8, 16, 32$, and each rule (the Most Efficient rule and the Least Efficient rule), we performed 500 independent simulations. Specifically, for each simulation, we started with the maximally floppy Miura-ori structure and added a quad planarity constraint at each step by randomly picking $k$ candidate facets and selecting one among them based on the prescribed selection rule. We then recorded the DOF change as the density of planarity constraints imposed $\rho$ increased from $0$ to $1$.

\begin{figure*}[t!]
    \centering
    \includegraphics[width=\linewidth]{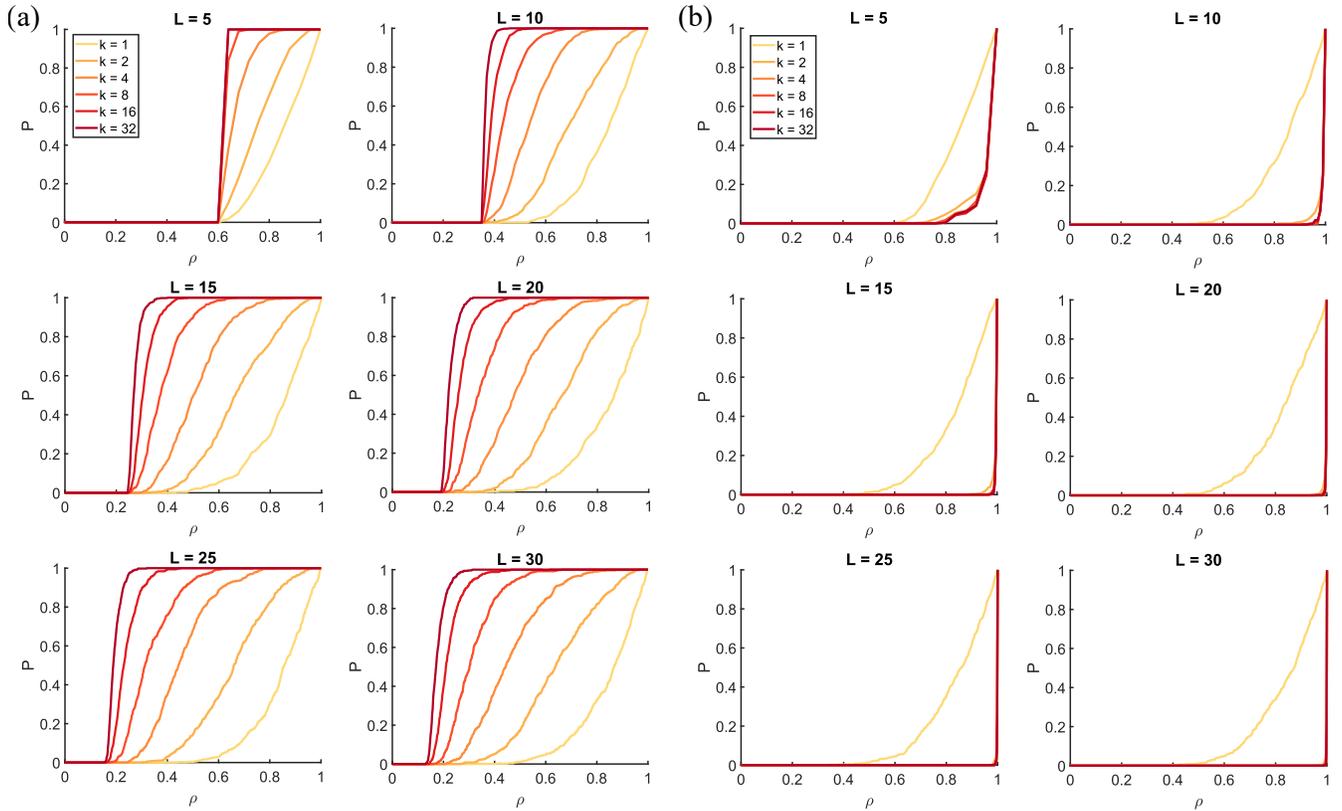}
\caption{\textbf{Rigidity percolation in origami under different selection rules for $L\times L$ Miura-ori structures.} (a) The Most Efficient selection rule. (b) The Least Efficient selection rule. For different problem size $L = 5, 10, 15, 20, 25, 30$ and different number of choices $k = 1, 2, 4, 8, 16, 32$, we calculated the probability $P$ of getting a 1-DOF structure at different planarity constraint density $\rho$ among the 500 simulations.}
    \label{fig:result_square_P}
\end{figure*}

As the DOF $d$ of an origami structure depends on the pattern sizes $m,n$ as shown in Eq.~\eqref{eqt:DOF}, here we consider the following normalized DOF $\widetilde{d}$ to facilitate the comparison across different pattern sizes:
\begin{equation}
    \widetilde{d} = \frac{d-1}{d_{\text{initial}}-1} = \frac{d-1}{4L-4}.
\end{equation}
It is easy to see that $\widetilde{d} \in [0,1]$. In Fig.~\ref{fig:result_square_dof_k}(a), we plot the value of $\widetilde{d}$ for all simulations under the Most Efficient selection rule for different values of $k$. For all $k$, the change of $\widetilde{d}$ shows a linear regime followed by a sublinear regime, which is consistent with the observation in the fully stochastic approach~\cite{chen2019rigidity}. Specifically, in the linear regime, we have $d = d_{\text{initial}} - t$ where $t$ is the number of planarity constraints imposed and $\widetilde{d} = (4L-3-t-1)/(4L-4) = 1 - t/(4L-4)$. In other words, the slope of $\widetilde{d}$ is given by $-1/(4L-4)$. Then, in the sublinear regime, $\widetilde{d}$ decreases gradually to 0 as $\rho$ increases. However, in contrast to the fully stochastic approach in~\cite{chen2019rigidity}, here we can see that by increasing the value of $k$, we can easily control the sharpness of the transition from the linear regime to the nonlinear regime. More specifically, increasing the value of $k$ effectively extends the linear regime and reduces the variation in $\widetilde{d}$ among the 500 simulations for each $L$.

By contrast, the change of the normalized DOF $\widetilde{d}$ under the Least Efficient selection rule exhibits a significantly different trend as shown in Fig.~\ref{fig:result_square_dof_k}(b). Specifically, while the change of $\widetilde{d}$ also shows a linear regime at small $\rho$ for all $k \geq 2$, in which adding every quad planarity constraint leads to a decrease in the DOF, there is subsequently a plateau regime in which $\widetilde{d}$ remains almost unchanged for a range of $\rho$. In this regime, the method preferentially selects the ``redundant'' quads for which adding the planarity constraint does not affect the DOF. As $\rho$ approaches 1, $\widetilde{d}$ enters another regime and decreases sharply to 0. Increasing the value of $k$ effectively extends the plateau regime and yields a sharper decrease in $\widetilde{d}$ near $\rho = 1$.

To conduct a more systematic analysis of the rigidity percolation transition, we consider the probability of getting a 1-DOF structure for each planarity constraint density $\rho$, defined as
\begin{equation}
    P(\rho) = \frac{\text{Number of 1-DOF structures at $\rho$}}{\text{Total number of simulations}}.
\end{equation}
For the Most Efficient selection rule, it can be observed in Fig.~\ref{fig:result_square_P}(a) that increasing the value of $k$ will lead to a sharper transition of $P$ from 0 to 1. For the Least Efficient selection rule, we can also see a notable difference between the transition behaviors for $k=1$ and $k>1$ in Fig.~\ref{fig:result_square_P}(b). 

\begin{figure*}[t]
    \centering
    \includegraphics[width=\linewidth]{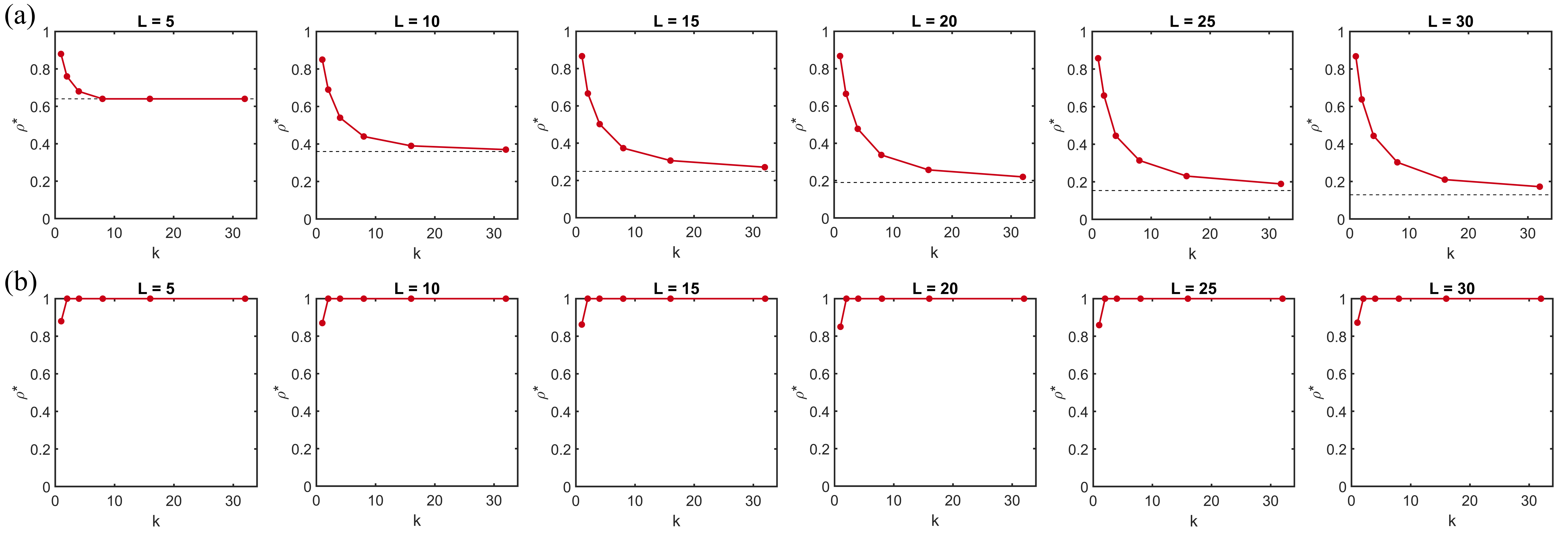}
    \caption{\textbf{The change in the critical transition $\rho^*$ with the number of choices $k$ for $L\times L$ Miura-ori structures.} (a)~The plots of the critical transition density $\rho^*$ obtained from our simulations under the Most Efficient selection rule against the number of choices $k = 1, 2, 4, 8, 16, 32$ for different pattern size. In each plot, the dotted line indicates the theoretical minimum density $\rho_{\min} = (4L-4)/L^2$ for getting a 1-DOF structure. (b) The plots of the critical transition density $\rho^*$ obtained from our simulations under the Least Efficient selection rule against the number of choices $k$ for different pattern size. Note that the theoretical maximum density for getting a 1-DOF structure is $\rho_{\max} = 1$.}
    \label{fig:result_square_k_vs_rho}
\end{figure*}

Now, note that a simple deterministic approach for constructing a 1-DOF origami structure using a minimal number of planarity constraints was proposed in~\cite{chen2019rigidity}. More precisely, it was shown that for even $L$, adding quad planarity constraints to all boundary facets in a $L\times L$ pattern is sufficient to make the structure 1-DOF. In this boundary-driven approach, the density of the planarity constraints added is
\begin{equation}
    \rho_b = \frac{\text{Number of boundary facets}}{\text{Total number of facets}} = \frac{4L-4}{L^2},
\end{equation}
which matches the theoretical minimum constraint density $\rho_{\min} = (2L+2L-4)/L^2 = (4L-4)/L^2$ in Eq.~\eqref{eqt:min_theoretical}. In other words, we have $\rho_{b} = \inf \{\rho: P = 1\}$. For $L = 10, 20, 30$, we have $\rho_{b} = 0.36, 0.19, 0.1289$ respectively. From the simulation results for $L = 10, 20, 30$ in Fig.~\ref{fig:result_square_P}(a), we can see that the transition from $P = 0$ to $P =1$ occurs at around these values of $\rho$. Also, it was shown in~\cite{chen2019rigidity} that for odd $L$, adding quad planarity constraints to all boundary facets will give a 2-DOF structure, where there will be an additional DOF involving the single center quad of the structure. In this boundary-driven approach, the planarity constraint density for getting a 1-DOF structure is
\begin{equation}
    \rho_b = \frac{\text{Number of boundary quads}+1}{\text{Number of quads}} = \frac{4L-3}{L^2}.
\end{equation}
Since the above $\rho_b$ differs from $\rho_{\min}$ in Eq.~\eqref{eqt:min_theoretical} by 1, it is natural to ask whether $\rho_b = (4L-3)/L^2$ is the actual minimum density required. For $L = 5, 15, 25$, we have $\rho_b = 0.68, 0.2533, 0.1552$ respectively. From the simulation results for $L = 5, 15, 25$ in Fig.~\ref{fig:result_square_P}(a), we can see that the transition from $P = 0$ to $P =1$ begins at comparable or even smaller values of $\rho$. For instance, there is a sharp transition at around $\rho = 0.6$ in the simulation results for $L = 5$, which is lower than $\rho_b = 0.68$. This shows that our power-of-choices strategy is more efficient than the intuitive boundary-driven approach in this case. Altogether, the above analysis shows that by introducing choices and applying the Most Efficient selection rule, we can effectively accelerate the phase transition, achieving performance that is comparable to, or even better than, that of intuitive deterministic design strategies.

As for the Least Efficient selection rule, note that as discussed previously, the theoretical maximum density is $\rho_{\max} = 1$ as one may delay the rigidity percolation transition by not enforcing the quad planarity of the corner facets until the end of the process. In our simulations, it is also easy to see that the rigidity percolation transition is significantly delayed by having $k > 1$. Interestingly, when compared to the Most Efficient selection rule for which we need a large value of $k$ to achieve a sharp transition in $P$, here in the Least Efficient selection rule, there is no notable difference between $k=2$ and other larger values $k=4,8,16,32$ as shown in Fig.~\ref{fig:result_square_P}(b).

\subsection{The optimal number of choices $k$}
As shown in the above analyses, having the ability to sample $k$ candidates and select one among them (the power of $k$ choices) dramatically changes the rigidity percolation behavior in origami. Ideally, using a large $k$ allows us to consider more candidates at each step and can lead to a better result. However, examining the effect of more candidates also increases the computational cost. It is natural to search for an optimal value of $k$ that gives a satisfactory performance while being as small as possible.

To quantify how the change in the number of choices $k$ affects the rigidity percolation transition, here we define the \emph{critical transition density} $\rho^*$ as the minimum $\rho$ with the probability of getting a 1-DOF structure $P \geq 1/2$ in our simulations. As shown in Fig.~\ref{fig:result_square_k_vs_rho}(a), under the Most Efficient selection rule, $\rho^*$ decreases and approaches the theoretical minimum density $\rho_{\min} = (4L-4)/L^2$ as the number of choices $k$ increases. As for the Least Efficient selection rule, from Fig.~\ref{fig:result_square_k_vs_rho}(b) we can see that $\rho^*$ increases rapidly from about 0.85 (for $k = 1$) to exactly 1 (for all $k \geq 2$).

\begin{figure}[t!]
    \centering
    \includegraphics[width=\linewidth]{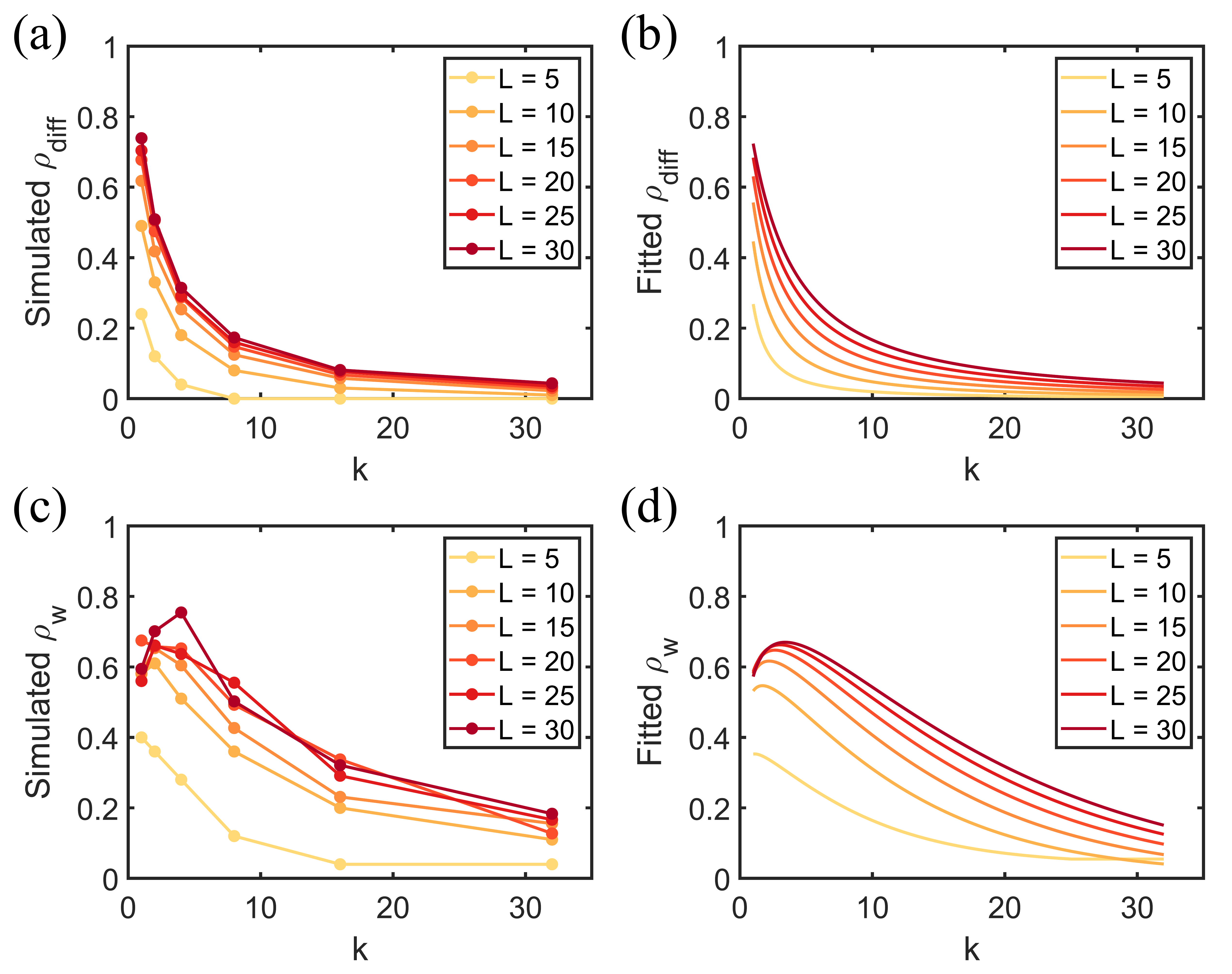}
    \caption{\textbf{Analysis of the critical percolation transition and the percolation transition width under the Most Efficient selection rule for $L\times L$ Miura-ori structures.} (a)~The values of $\rho_{\text{diff}} = \rho^* - \rho_{\min}$ from the simulation results. Here, each data point represents the result obtained from a set of 500 simulations with a specific pair of parameters $(L, k)$ (where $L = 5, 10, 15, 20, 25, 30$ and $k = 1, 2, 4, 8, 16, 32$). (b)~The fitted values of $\rho_{\text{diff}}$ by Eq.~\eqref{eqt:law_diff}. (c)~The percolation transition width $\rho_{w} = \rho_1 - \rho^0$ from the simulation results. (d)~The fitted values of $\rho_{w}$ by Eq.~\eqref{eqt:law_width}.}
    \label{fig:analysis_square}
\end{figure}

From the above results, it is clear that $k=2$ is an optimal choice for delaying the rigidity percolation transition, as it achieves the same effect (with $\rho^* = 1$) as other larger $k$, while requiring less computation. As for accelerating the rigidity percolation transition, we further consider the difference between the critical transition density and the theoretical minimum density, i.e., $\rho_{\text{diff}} = \rho^* - \rho_{\min}$. As shown in Fig.~\ref{fig:analysis_square}(a), the values of $\rho_{\text{diff}}$ from the simulation results decrease rapidly with $k$. Also, it can be observed that for any fixed $k$, the value of $\rho_{\text{diff}}$ increases gradually with the pattern size $L$. Lastly, note that $\rho_{\text{diff}}$ should be within the range $[0,1]$ for any $k, L$. Therefore, we consider the following simple relationship between $\rho_{\text{diff}}$, $k$ and $L$: $\rho_{\text{diff}} = a(1+b(k/L))^{-c}$, where $a,b,c$ are parameters. By fitting the simulation results using the above model, we obtain:
\begin{equation}\label{eqt:law_diff}
    \rho_{\text{diff}}(L,k) \approx \left(1+\frac{7.8 k}{L}\right)^{-1.4}.
\end{equation}
It can be observed in Fig.~\ref{fig:analysis_square}(b) that the fitted model matches the simulation results very well. The simple scaling law above provides an efficient way to determine a suitable value of $k$ to achieve a target accuracy for the sharp transition. For instance, in order to have an accuracy of $\rho_{\text{diff}} = \rho^* - \rho_{\min} \approx 0.05$, we need $k \approx 0.96 L$. Also, to achieve $\rho_{\text{diff}} \approx 0.01$, we need $k \approx 3.3L$.

In the above analysis, we considered the optimality of the number of choices $k$ in terms of the difference $\rho_{\text{diff}}$. Alternatively, one may assess the optimality of $k$ by considering the sharpness of the rigidity percolation transition. Specifically, we define the \emph{percolation transition width} as $\rho_w = \rho_1 - \rho^0$, where $\rho_1$ is the minimum $\rho$ with $P = 1$ and $\rho^0$ is the maximum $\rho$ with $P = 0$ in our simulations. It is natural to ask whether increasing $k$ can lead to a decrease in $\rho_w$ and a sharper transition. Surprisingly, the transition width $\rho_w$ generally first increases as we increase the number of choices from $k=1$ to some small $k$ (see Fig.~\ref{fig:analysis_square}(c)). Then, the transition width eventually decreases as we further increase $k$. A possible explanation of this counter-intuitive observation is that while introducing choices generally allows both $\rho_1$ and $\rho_0$ to move towards the theoretical transition value $\rho_{\min}$, they do not necessarily change at the same rate. In Appendix~\ref{sec:appendix_additional}, we further provide the statistics of different values of $\rho^a$ (defined as the maximum $\rho$ with $P = a$) and $\rho_b$ (defined as the minimum $\rho$ with $P = b$), from which we can see a similar trend for different choices of the transition interval. In other words, increasing the sharpness of the rigidity percolation transition width requires a relatively large $k$, while simply increasing from $k=1$ to $k=2$ may lead to an adverse effect. 

The above observations motivate us to consider fitting $\rho_w$ by separately fitting $\rho^0$ and $\rho_1$. In particular, note that as $k$ increases, we should have $\rho^0 \approx \rho_{\min}- 1/L^2$. Hence, we consider $\rho^0 \approx  (\rho_{\min}- 1/L^2) + (1-(\rho_{\min}- 1/L^2)) \exp(-a_0 k^{b_0}/L^{c_0})$, where $a_0,b_0,c_0$ are parameters. Similarly, note that as $k$ increases, we should have $\rho_1 \approx \rho_{\min}$. Hence, we consider $\rho_1 \approx  \rho_{\min} + (1-\rho_{\min}) \exp(-a_1 k^{b_1}/L^{c_1})$, where $a_1, b_1, c_1$ are parameters. We obtain the following fitted models for $\rho^0$ and $\rho_1$ (see also Appendix~\ref{sec:appendix_additional}):
\begin{equation}\label{eqt:fit_rho0}
    \rho^0  \approx \frac{4L-5}{L^2} + \left(1- \frac{4L-5}{L^2}\right) \exp\left(-6.3 \sqrt{\frac{k}{L}}\right),
\end{equation}
\begin{equation}\label{eqt:fit_rho1}
    \rho_1  \approx \frac{4L-4}{L^2} + \left(1- \frac{4L-4}{L^2}\right) \exp\left(-0.15 \frac{k^{1.2}}{\sqrt{L}}\right),
\end{equation}
and hence we obtain the following fitted model for $\rho_{w}$:
\begin{equation} \label{eqt:law_width}
\begin{split}
    \rho_{w}(L,k) \approx  \frac{1}{L^2} &+ \left(1- \frac{4L-4}{L^2}\right) \exp\left(-0.15 \frac{k^{1.2}}{\sqrt{L}}\right) \\
    &- \left(1- \frac{4L-5}{L^2}\right) \exp\left(-6.3 \sqrt{\frac{k}{L}}\right).
\end{split}
\end{equation}
As shown in Fig.~\ref{fig:analysis_square}(d), the fitted model qualitatively matches the overall trend of the simulated $\rho_w$. Notably, it shows a similar initial increase for small $k$ followed by a gradual decrease. Using Eq.~\eqref{eqt:law_width}, we have an alternative way to determine the number of choices $k$ needed for achieving a target sharpness of the rigidity percolation transition. For instance, to achieve $\rho_w \approx 0.1$ for the pattern size $L \times L = 30 \times 30$, we need $k \approx 39 = 1.3 L$. To achieve $\rho_w \approx 0.05$, we need $k \approx 48 = 1.6L$.

\begin{figure}[t]
    \centering
    \includegraphics[width=\linewidth]{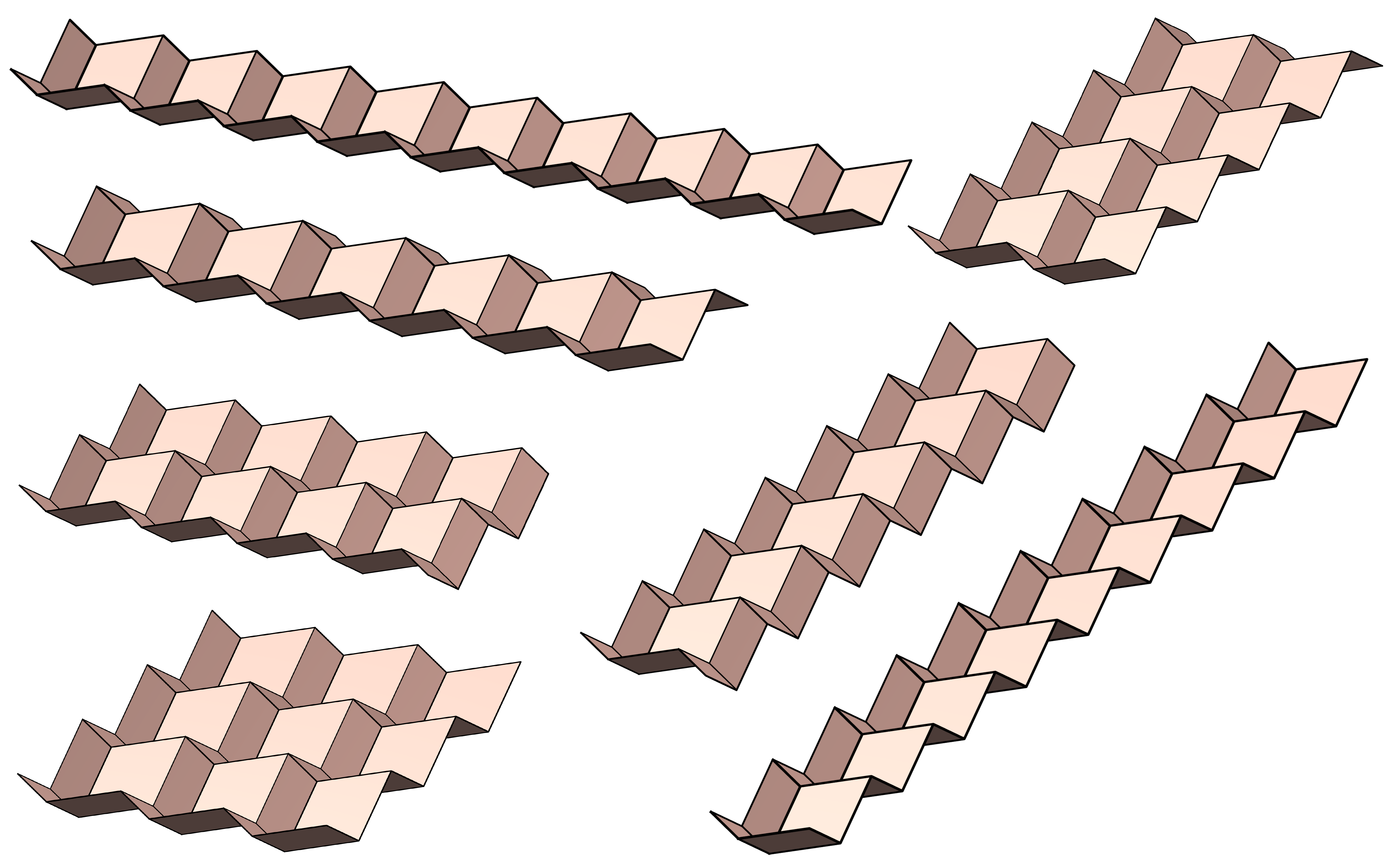}
    \caption{\textbf{General $m\times n$ Miura-ori structures.} For Miura-ori structures with $S = 36$ facets, all dimensions including $2\times 18$, $3 \times 12$, $4\times 9$, $6\times 6$, $9 \times 4$, $12\times 3$ and $18\times 2$ are considered.}
    \label{fig:size_rect}
\end{figure}

\subsection{The general rectangular case}
After performing the analyses on the square case where the Miura-ori structures have the same number of rows and columns of quads ($m=n=L$), we consider the general rectangular case with $m$ and $n$ not necessarily the same. In particular, it is natural to study whether the explosive rigidity percolation transition is affected not only by the pattern size ($mn$) and the number of choices ($k$) but also by the ratio between the two dimensions ($m/n$).

Note that for $m=1$ or $n=1$, the resulting origami structure is just a strip of quads, and hence the DOF is always greater than 1, even if all quads are planar. Therefore, in the following, we consider integers $m,n \geq 2$. In particular, for a given positive integer $S$ representing the pattern size, we consider expressing $S = mn$ for all possible combinations of $m,n\geq2$ (see Fig.~\ref{fig:size_rect} for an illustration). Analogous to the previous case, 500 simulations were performed for each combination of $(m,n)$ and each number of choices $k$, and the probability of obtaining a 1-DOF structure $P$ was calculated at various quad planarity constraint densities $\rho$. The difference between the critical transition density, $\rho_{\text{diff}} = \rho^* - \rho_{\min}$, and the percolation transition width, $\rho_{w} = \rho_1 - \rho^0$, can then be analyzed as in the square case.

\begin{figure}[t]
    \centering
    \includegraphics[width=\linewidth]{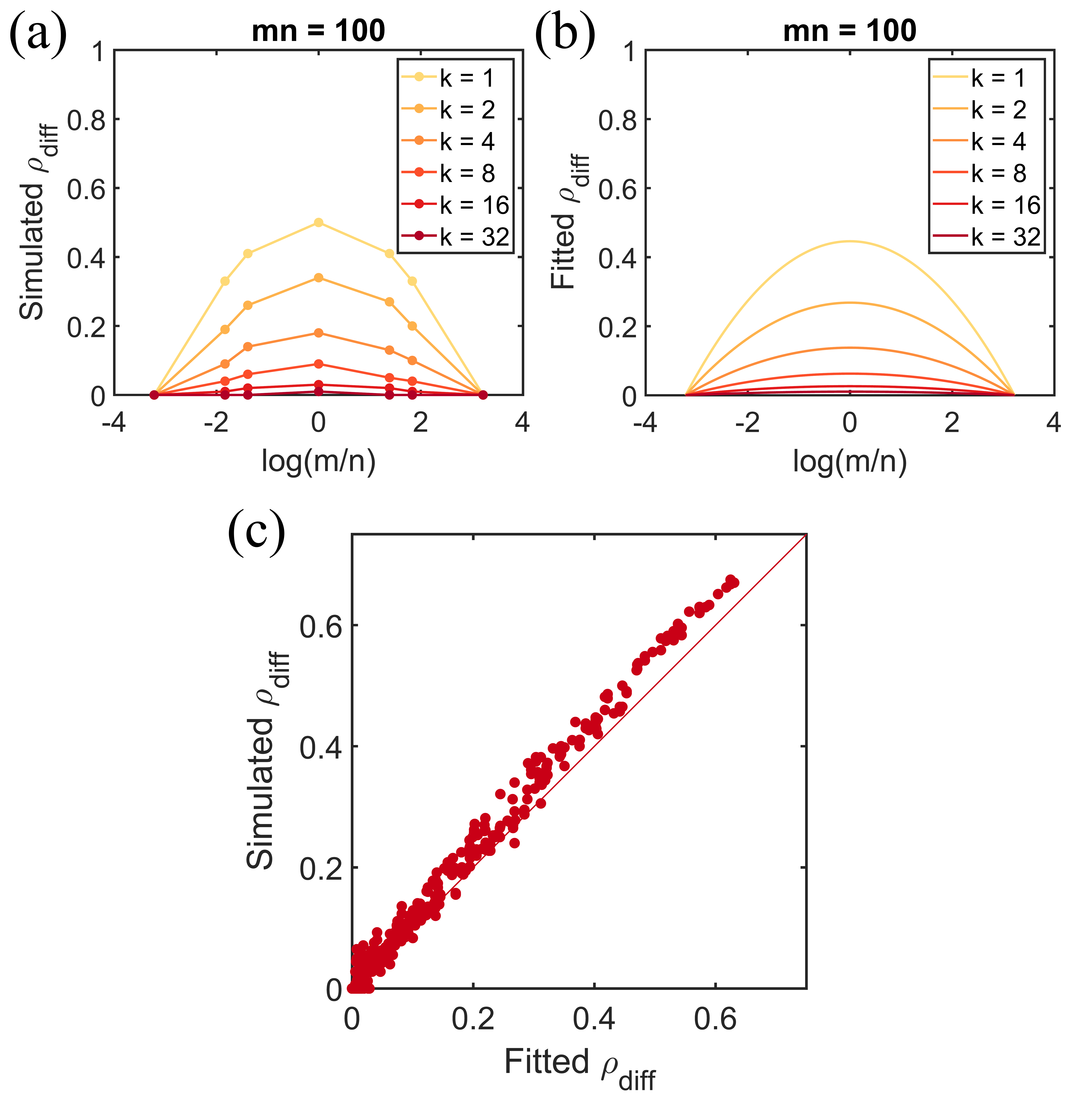}
    \caption{\textbf{Analysis of the difference between the critical transition density and the theoretical minimum density $\rho_{\text{diff}} = \rho^* - \rho_{\min}$ for general $m\times n$ Miura-ori structures.} (a) The simulated $\rho_{\text{diff}}$ against $\log(m/n)$ for $mn = 100$ obtained based on the Most Efficient selection rule. Here, each data point represents the result obtained from a set of 500 simulations in a specific setup $(m,n,k)$ (where $mn = 100$ with $m,n \geq 2$ and $k = 1, 2, 4, 8, 16, 32$). (b) The fitted $\rho_{\text{diff}}$ for $mn = 100$ obtained using our model. (c) The plot of the simulated $\rho_{\text{diff}}$ against the fitted $\rho_{\text{diff}}$ for different combinations of $(m,n,k)$ with $mn = 25, 36, 49, 64, \dots, 400$ and $k = 1,2,4,8,16,32$. The red line represents $y = x$.}
    \label{fig:analysis_rect_rdiff}
\end{figure}

In Fig.~\ref{fig:analysis_rect_rdiff}(a), we consider all combinations $mn = 100$ with $m, n \geq 2$ and study the relationship between $\rho_{\text{diff}}$ and $\log(m/n)$ for different $k$. It can be observed for any fixed $k$, the transition width $\rho_w$ is higher as $\log (m/n)$ is closer to 0, i.e. the square case $m = n$. By contrast, as the ratio between $m$ and $n$ becomes more extreme, the transition width decreases significantly. In particular, the transition width is always at the minimum when either $m = 2$ or $n = 2$. This can be explained by the fact that for $m = 2$ or $n = 2$, all facets of the origami structure are boundary facets. Then, putting either $m=2$ or $n=2$ into Eq.~\eqref{eqt:min_theoretical}, we have 
\begin{equation}
    \rho_{\min} = \frac{2(2)+2(mn/2)-4}{2(mn/2)} = \frac{mn}{mn} = 1.
\end{equation}
In other words, turning the initial floppy structure into a 1-DOF structure requires explicitly adding the planarity constraints to all facets. Consequently, the smallest $\rho_{\text{diff}}$ is always achieved in this case. Besides, we can easily see that increasing the number of choices $k$ leads to a decrease in $\rho_{\text{diff}}$. Now, from the symmetry of $\rho_{\text{diff}}$ about $\log(m/n) = 0$, it is natural to ask whether one can approximate $\rho_{\text{diff}}$ using a simple polynomial in $\log(m/n)$ together with the information at the peak of the curves. This motivates us to consider the following model for $\rho_{\text{diff}}$ for general $(m,n,k)$:
\begin{equation} \label{eqt:law_diff_rect}
    \rho_{\text{diff}}(m,n,k) \approx \left(1+\frac{7.8 k}{\sqrt{mn}}\right)^{-1.4} \left(1 - \left(\frac{\log (m/n)}{\log (mn/4)}\right)^2\right).
\end{equation}
Here, the factor $\left(1+{7.8 k}/\sqrt{mn}\right)^{-1.4}$ follows from the square case in Eq.~\eqref{eqt:law_diff}. It is easy to see that if $m=n$, Eq.~\eqref{eqt:law_diff_rect} becomes identical to Eq.~\eqref{eqt:law_diff}. 
Also, if $n=2$, we have 
\begin{equation}
\left(\frac{\log (m/n)}{\log (mn/4)}\right)^2 = \left(\frac{\log ((mn/2)/2)}{\log (mn/4)}\right)^2 = 1
\end{equation}
and hence Eq.~\eqref{eqt:law_diff_rect} becomes 0. Similarly, if $m = 2$, we have $((\log(m/n))/(\log(mn/4)))^2 = (-1)^2 = 1$. This shows that Eq.~\eqref{eqt:law_diff_rect} matches the expected $\rho_{\text{diff}}$ at the peak and the two endpoints. As shown in Fig.~\ref{fig:analysis_rect_rdiff}(b), the fitted $\rho_{\text{diff}}$ for $mn = 100$ matches the simulation results very well. To further verify this relationship, we performed additional simulations for $mn = 25, 36, 49, 64, \dots, 400$, with all possible $m,n \geq 2$ and number of choices $k = 1,2,4,8,16,32$ considered (500 simulations for each combination of $(m,n,k)$, over 260,000 simulations in total). As shown in Fig.~\ref{fig:analysis_rect_rdiff}(c), the simulated $\rho_{\text{diff}}$ and the fitted $\rho_{\text{diff}}$ are highly consistent (see Appendix~\ref{sec:appendix_mn} for more results).

\begin{figure}[t]
    \centering
    \includegraphics[width=\linewidth]{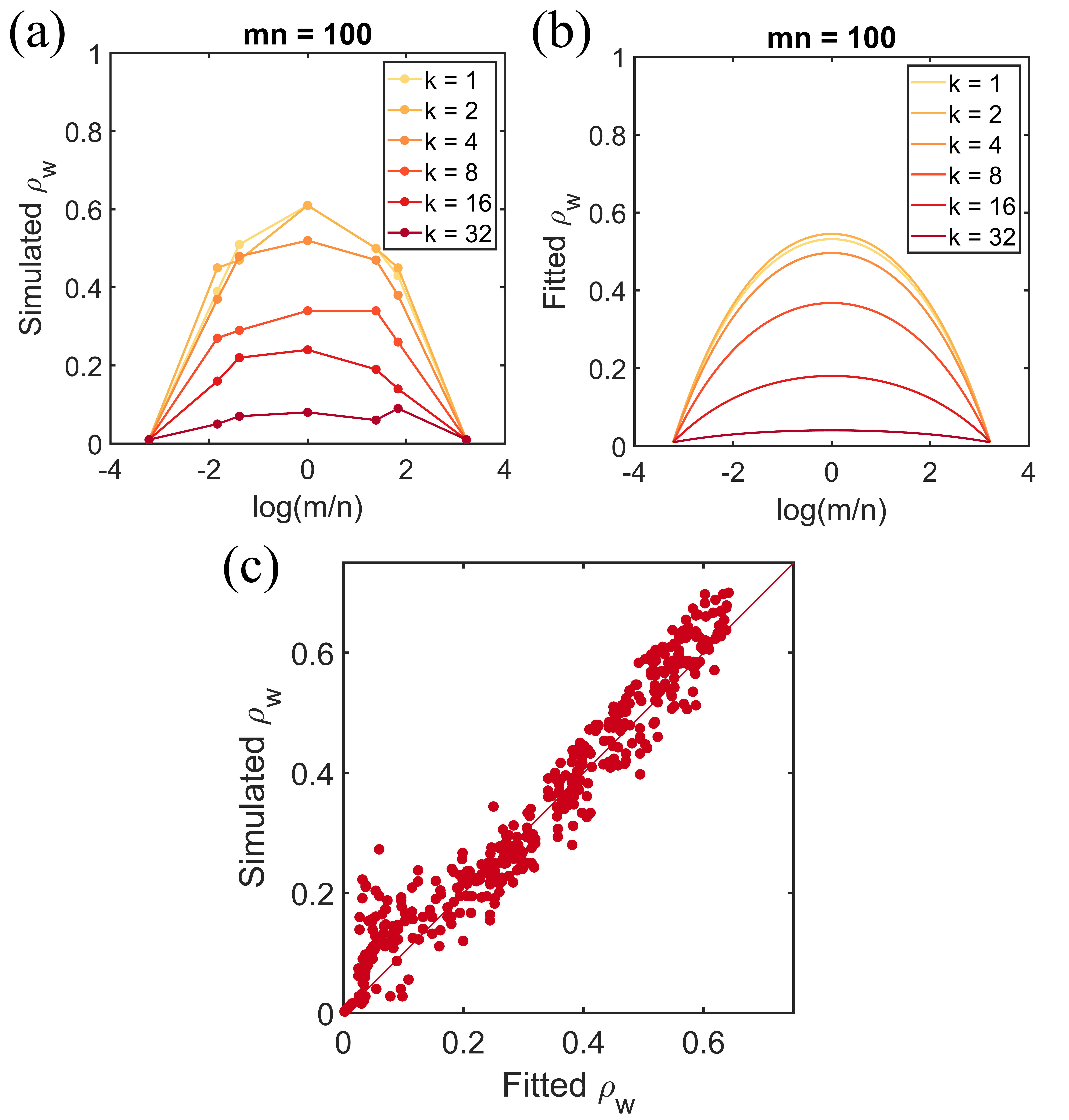}
    \caption{\textbf{Analysis of the rigidity percolation transition width $\rho_w$ for general $m\times n$ Miura-ori structures.} (a) The simulated $\rho_{w}$ against $\log(m/n)$ for $mn = 100$ obtained based on the Most Efficient selection rule. Here, each data point represents the result obtained from a set of 500 simulations in a specific setup $(m,n,k)$ (where $mn = 100$ with $m,n \geq 2$ and $k = 1, 2, 4, 8, 16, 32$). (b) The fitted $\rho_w$ for $mn = 100$ obtained using our model. (c) The plot of the simulated $\rho_w$ against the fitted $\rho_w$ for different combinations of $(m,n,k)$ with $mn = 25, 36, 49, 64, \dots, 400$ and $k = 1,2,4,8,16,32$. The red line represents $y = x$.}
    \label{fig:analysis_rect}
\end{figure}

As for the rigidity percolation transition width $\rho_w$, in Fig.~\ref{fig:analysis_rect}(a) we again plot the simulated $\rho_w$ for $mn = 100$, from which we see a roughly symmetric trend. Moreover, comparing the results for different $k$, it can be observed that increasing the number of choices from $k = 1$ to some small $k$ leads to a slight increase in the transition width $\rho_w$, which is consistent with our observation in the square case ($L \times L$) in the previous section. Then, as $k$ further increases, the transition width gradually decreases. To quantitatively describe the above observations, here we extend Eq.~\eqref{eqt:law_width} by using the general formula in Eq.~\eqref{eqt:min_theoretical} for $\rho_{\text{min}}$ and replacing $L$ with $\sqrt{mn}$. We obtain the following approximation formula of $\rho_w$ for the general rectangular case:
\begin{equation} \label{eqt:law_width_rect}
\begin{split}
        & \rho_w (m,n,k) \\
        \approx & \frac{1}{mn} + \left(1- \frac{2m+2n-4}{mn}\right) \exp\left(-0.15 \frac{k^{1.2}}{(mn)^{1/4}}\right) \\
    &- \left(1- \frac{2m+2n-5}{mn}\right) \exp\left(-6.3 \frac{\sqrt{k}}{(mn)^{1/4}}\right).
\end{split}
\end{equation}
As shown in Fig.~\ref{fig:analysis_rect}(b), the fitted $\rho_w$ for $mn = 100$ matches the simulation results very well. Analogous to the above analysis for $\rho_{\text{diff}}$, in Fig.~\ref{fig:analysis_rect}(c) we further compare the simulated $\rho_w$ and the fitted $\rho_w$ for all combinations $(m,n,k)$ with $mn = 36, 49, 64, \dots, 400$ and $k = 1, 2, 4, 8, 16, 32$. It can be observed that the simulated and fitted values are highly consistent (see Appendix~\ref{sec:appendix_mn} for more results). 

Altogether, our analysis shows that the phenomena we observe in the square case can be naturally extended to the general rectangular case via a simple modification in the fitted models.

\section{Discussion}
In this work, we have studied the rigidity percolation in origami structures and demonstrated how \emph{the power of $k$ choices} together with some simple selection rules can lead to explosive rigidity percolation transition in origami. Moreover, we have derived simple formulas that relate $k$ with the origami pattern size and the sharpness of the rigidity percolation transition. More broadly, our work suggests that we can easily control the rigidity of mechanical metamaterials using a fusion of deterministic and stochastic approaches, thereby shedding light on the design of mechanical metamaterials with potential applications to mechanical information storage and soft robotics.

While we have focused on the Miura-ori pattern in this work, it is noteworthy that our analyses of infinitesimal rigidity rely only on the edge, diagonal, and facet planarity constraints and hence are also applicable to other origami patterns. In our future work, we plan to study a wider class of origami structures to analyze how the structural arrangements of the origami facets affect the rigidity percolation transition. Another natural next step is to extend our study to the control of two- or three-dimensional structural assemblies~\cite{lubbers2019excess,overvelde2017rational,choi2020control}.

\bibliographystyle{ieeetr}
\bibliography{reference}

\clearpage

\centerline{\large\textbf{Supplementary Information}}
\appendix
\renewcommand\thefigure{S\arabic{figure}}    
\setcounter{figure}{0}
\renewcommand\thetable{S\arabic{table}}    
\setcounter{table}{0}


\section{Construction of the infinitesimal rigidity matrix}
\label{sec:appendix_matrix}

In the main text, we provided the explicit formulas for the partial derivatives of each edge constraint in the infinitesimal rigidity matrix $A$. Here, we give the explicit formulas for the partial derivatives of the other constraints in $A$.

For each diagonal (no-shear) constraint, suppose $\mathbf{v}_{i_3} = (x_{i_3}, y_{i_3}, z_{i_3})$ and $\mathbf{v}_{i_1} = (x_{i_1}, y_{i_1}, z_{i_1})$. We can explicitly derive the partial derivatives of $g_d$ as follows:
\begin{align}
    \frac{\partial g_d}{\partial x_{i_3}} = -\frac{\partial g_d}{\partial x_{i_1}} = 2(x_{i_3}- x_{i_1}),\\
    \frac{\partial g_d}{\partial y_{i_3}} = -\frac{\partial g_d}{\partial y_{j_1}} = 2(y_{i_3}- y_{i_1}),\\
    \frac{\partial g_d}{\partial z_{i_3} } = -\frac{\partial g_d}{\partial z_{i_1}} = 2(z_{i_3}- z_{i_1}),
\end{align}
and the partial derivatives of $g_d$ with respect to all other variables are 0. This shows that each row of $A$ associated with a diagonal constraint has at most 6 non-zero entries. 

Lastly, for each quad planarity constraint added to the system, suppose $\mathbf{v}_{i_j} = (x_{i_j}, y_{i_j}, z_{i_j})$ where $j = 1,2,3,4$. We can explicitly derive the partial derivatives of $g_p$ as follows:
\begin{equation}
    \begin{aligned}
        \frac{\partial g_p}{\partial x_{i_1}} = & -(y_{i_2} - y_{i_1})(z_{i_4} - z_{i_1}) 
        + (y_{i_4} - y_{i_3})(z_{i_2} - z_{i_1}) \\
        & - (y_{i_3} - y_{i_1})(z_{i_2} - z_{i_1}) 
        + (y_{i_3} - y_{i_1})(z_{i_4} - z_{i_1}) \\
        & - (y_{i_4} - y_{i_1})(z_{i_3} - z_{i_1}) 
        + (y_{i_2} - y_{i_1})(z_{i_3} - z_{i_1}), 
    \end{aligned}
\end{equation}
\begin{equation}
    \begin{aligned}
        \frac{\partial g_p}{\partial y_{i_1}} = & -(x_{i_3} - x_{i_1})(z_{i_4} - z_{i_1}) 
        -(x_{i_4} - x_{i_1})(z_{i_2} - z_{i_1}) \\
        & - (x_{i_2} - x_{i_1})(z_{i_3} - z_{i_1}) 
        + (x_{i_2} - x_{i_1})(z_{i_4} - z_{i_1}) \\
        & + (x_{i_4} - x_{i_1})(z_{i_3} - z_{i_1}), 
    \end{aligned}
\end{equation}
\begin{equation}
    \begin{aligned}
        \frac{\partial g_p}{\partial z_{i_1}} = & -(x_{i_3} - x_{i_1})(y_{i_2} - y_{i_1}) 
        + (x_{i_3} - x_{i_1})(y_{i_4} - y_{i_3}) \\
        & - (x_{i_4} - x_{i_1})(y_{i_3} - y_{i_1}) 
        + (x_{i_2} - x_{i_1})(y_{i_3} - z_{i_1}) \\
        & - (x_{i_2} - x_{i_1})(y_{i_4} - y_{i_1}) 
        + (x_{i_4} - x_{i_1})(y_{i_2} - y_{i_1}), 
    \end{aligned}
\end{equation}
\begin{equation}
    \begin{aligned}
        \frac{\partial g_p}{\partial x_{i_2}} = & -(y_{i_3} - y_{i_1})(z_{i_4} - z_{i_1}) 
        + (y_{i_4} - y_{i_1})(z_{i_3} - z_{i_1}),
    \end{aligned}
\end{equation}
\begin{equation}
    \begin{aligned}
        \frac{\partial g_p}{\partial y_{i_2}} = & (x_{i_3} - x_{i_1})(z_{i_4} - z_{i_1}) 
        - (x_{i_4} - x_{i_1})(z_{i_3} - z_{i_1}),
    \end{aligned}
\end{equation}
\begin{equation}
    \begin{aligned}
        \frac{\partial g_p}{\partial z_{i_2}} = & (x_{i_3} - x_{i_1})(y_{i_4} - y_{i_3}) 
        - (y_{i_3} - y_{i_1})(x_{i_4} - x_{i_1}),
    \end{aligned}
\end{equation}
\begin{equation}
    \begin{aligned}
        \frac{\partial g_p}{\partial x_{i_3}} = & - (y_{i_2} - y_{i_1})(z_{i_4} - z_{i_1}) 
        - (y_{i_4} - y_{i_3})(z_{i_2} - z_{i_1}),
    \end{aligned}
\end{equation}
\begin{equation}
    \begin{aligned}
        \frac{\partial g_p}{\partial y_{i_3}} = & (x_{i_3} - x_{i_1})(z_{i_2} - z_{i_1}) 
        + (x_{i_4} - x_{i_1})(z_{i_2}- z_{i_1})\\
        & -  (x_{i_2} - x_{i_1})(z_{i_4} - z_{i_1}),
    \end{aligned}
\end{equation}
\begin{equation}
    \begin{aligned}
        \frac{\partial g_p}{\partial z_{i_3}} = & (x_{i_2} - x_{i_1})(y_{i_4} - y_{i_3}) 
        - (y_{i_2} - y_{i_1})(x_{i_4} - x_{i_1}),
    \end{aligned}
\end{equation}
\begin{equation}
    \begin{aligned}
        \frac{\partial g_p}{\partial x_{i_4}} = & (y_{i_3} - y_{i_1})(z_{i_2} - z_{i_1}) - (y_{i_2} - y_{i_1})(z_{i_3} - z_{i_1}),
    \end{aligned}
\end{equation}
\begin{equation}
    \begin{aligned}
        \frac{\partial g_p}{\partial y_{i_4}} = & -(x_{i_3} - x_{i_1})(z_{i_2} - z_{i_1}) + (x_{i_2} - x_{i_1})(z_{i_3} - z_{i_1}),
    \end{aligned}
\end{equation}
\begin{equation}
    \begin{aligned}
        \frac{\partial g_p}{\partial z_{i_4}} = & (x_{i_3} - x_{i_1})(y_{i_2} - y_{i_1}) - (x_{i_2} - x_{i_1})(y_{i_3} - y_{i_1}),
    \end{aligned}
\end{equation}
and the partial derivatives of $g_p$ with respect to all other variables are 0. This shows that each row of $A$ associated with a quad planarity constraint has at most 12 non-zero entries. 

Altogether, the infinitesimal rigidity matrix $A$ is a sparse matrix and all entries of it can be explicitly expressed in terms of the vertex coordinates of the origami structure.

\begin{figure*}[t!]
    \centering
    \includegraphics[width=0.9\linewidth]{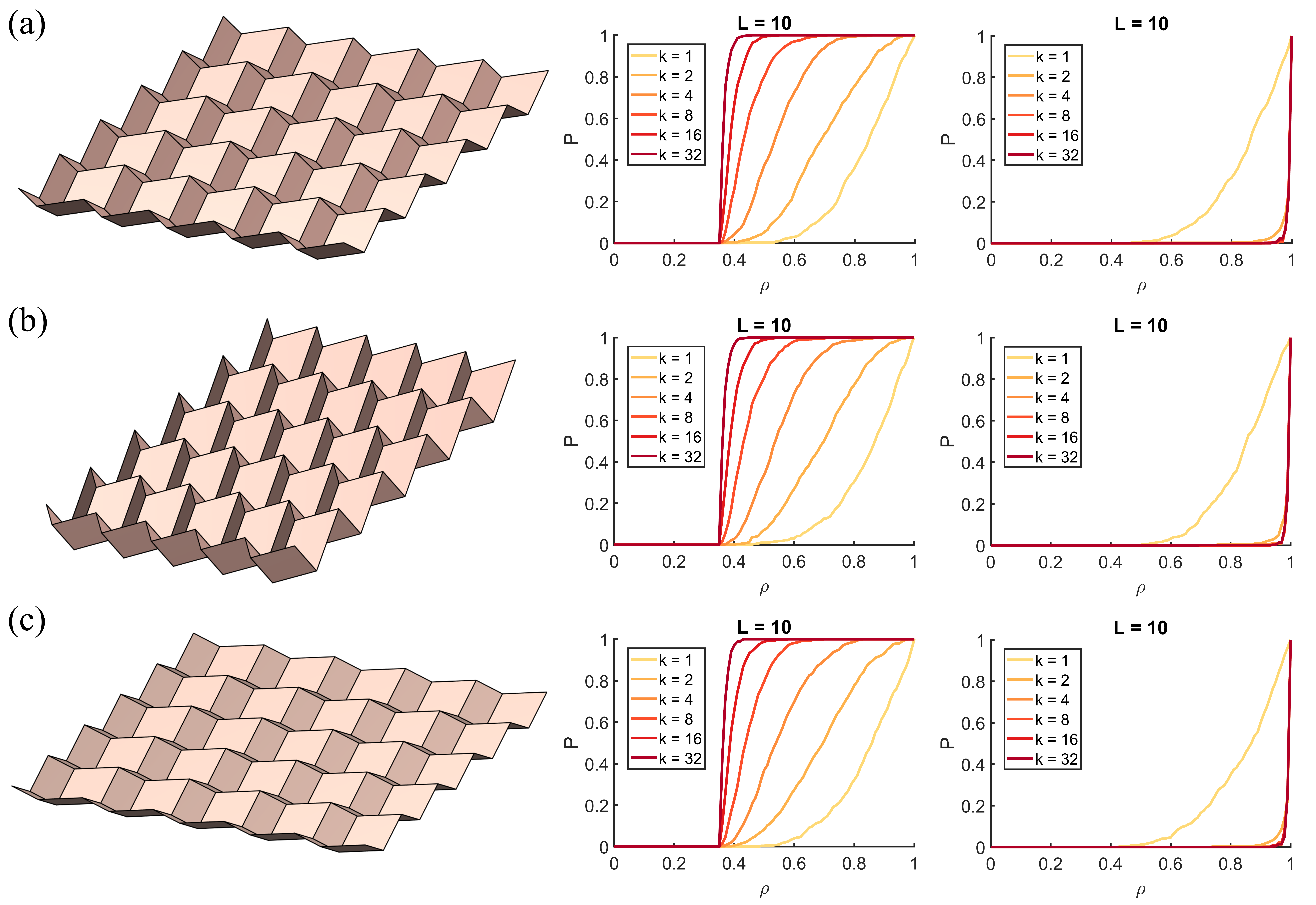}
    \caption{\textbf{Comparing the explosive rigidity percolation in origami with different geometric parameters.} (a) The results for $\gamma = \pi/4$ and $\theta = \cos^{-1} \sqrt{2/3}$. (b) The results for $\gamma = \pi/3$ and $\theta = \pi/3$. (c) The results for $\gamma = \pi/6$ and $\theta = \pi/6$. For each set of geometric parameters, we consider a $10 \times 10$ Miura-ori structure (left), the rigidity percolation simulation result based on the Most Efficient selection rule with different number of choices $k$, and the simulation result based on the Least Efficient rule (right). Here, $\rho$ is the density of the planarity constraints explicitly imposed and $P$ is the probability of getting a 1-DOF structure.}
    \label{fig:SI_comparison_geometry}
\end{figure*}

\section{Changing the geometry of the origami structure}
\label{sec:appendix_geometry}

It is noteworthy that the calculation of the infinitesimal rigidity matrix $A$ involves the vertex coordinates of the Miura-ori structure. As discussed in~\cite{chen2019rigidity}, changing the geometric parameters of the Miura-ori structure, such as the angles $\gamma$ and $\theta$, does not affect the rigidity percolation. It is natural to further ask whether the explosive percolation transition will be affected by the geometry of the Miura-ori structure. To address this question, we changed the origami geometry and repeated the simulations with the two prescribed selection rules. 

In Fig.~\ref{fig:SI_comparison_geometry}(a), we show the geometry of Miura-ori structure used in the simulations in the main text (with the angle parameters $\gamma = \pi/4$ and $\theta = \cos^{-1} \sqrt{2/3}$) and the simulation results for $m=n=L =10$ and $k=1,2,4,8,16,32$ based on the two selection rules. Then, we changed the angle parameters to be $\gamma = \pi/3$ and $\theta = \pi/3$ and repeated the simulations (500 simulations for each $L$ and each $k$ as in the main text). As shown in Fig.~\ref{fig:SI_comparison_geometry}(b), the geometry of the Miura-ori structure is significantly different from the original one. Nevertheless, for both the Most Efficient selection rule and the Least Efficient selection rule, the simulation results are highly consistent with the ones obtained under the original setup. Specifically, it can be observed that increasing the number of choices $k$ from $1$ to $32$ gives a highly similar trend in the increase of the sharpness of the transition of the probability $P$ of getting a 1-DOF structure. In Fig.~\ref{fig:SI_comparison_geometry}(c), we further considered another set of angle parameters $\gamma = \pi/6$ and $\theta = \pi/6$ for the Miura-ori geometry and repeated the simulations. Again, one can see that the simulation results are highly consistent with the above-mentioned ones. 

From the above additional experiments, we conclude that the explosive rigidity percolation transition is independent of the geometry of the Miura-ori structure.

\begin{table*}[t]
    \centering
    \begin{tabular}{|c||c||c|c|c||c|c|c||c|c|c|} \hline
        Pattern size ($L\times L$) & \# choices ($k$) & $\rho^0$ & $\rho_1$ & $\rho_1-\rho^0$ & $\rho^{0.1}$ & $\rho_{0.9}$ & $\rho_{0.9}-\rho^{0.1}$ & $\rho^{0.25}$ & $\rho_{0.75}$ & $\rho_{0.75}-\rho^{0.25}$ \\ \hline
         \multirow{6}{*}{$5\times 5$}  & 1 & 0.6000 & 1.0000 & 0.4000 & 0.6800 & 1.0000 & 0.3200 & 0.7600 & 0.9600 & 0.2000\\
& 2 & 0.6000 & 0.9600 & 0.3600 & 0.6400 & 0.8800 & 0.2400 & 0.6800 & 0.8400 & 0.1600\\
& 4 & 0.6000 & 0.8800 & 0.2800 & 0.6000 & 0.7600 & 0.1600 & 0.6000 & 0.7200 & 0.1200\\
& 8 & 0.6000 & 0.7200 & 0.1200 & 0.6000 & 0.6800 & 0.0800 & 0.6000 & 0.6400 & 0.0400\\
& 16 & 0.6000 & 0.6400 & 0.0400 & 0.6000 & 0.6400 & 0.0400 & 0.6000 & 0.6400 & 0.0400\\
& 32 & 0.6000 & 0.6400 & 0.0400 & 0.6000 & 0.6400 & 0.0400 & 0.6000 & 0.6400 & 0.0400\\ \hline
 \multirow{6}{*}{$10\times 10$}  & 1 & 0.4200 & 1.0000 & 0.5800 & 0.6700 & 0.9700 & 0.3000 & 0.7500 & 0.9300 & 0.1800\\
& 2 & 0.3600 & 0.9700 & 0.6100 & 0.5300 & 0.8700 & 0.3400 & 0.6000 & 0.8000 & 0.2000\\
& 4 & 0.3500 & 0.8600 & 0.5100 & 0.4300 & 0.6800 & 0.2500 & 0.4700 & 0.6100 & 0.1400\\
& 8 & 0.3500 & 0.7100 & 0.3600 & 0.3700 & 0.5400 & 0.1700 & 0.3900 & 0.4900 & 0.1000\\
& 16 & 0.3500 & 0.5500 & 0.2000 & 0.3500 & 0.4400 & 0.0900 & 0.3700 & 0.4100 & 0.0400\\
& 32 & 0.3500 & 0.4600 & 0.1100 & 0.3500 & 0.3900 & 0.0400 & 0.3500 & 0.3700 & 0.0200\\ \hline
 \multirow{6}{*}{$15\times 15$}  & 1 & 0.4178 & 1.0000 & 0.5822 & 0.6756 & 0.9778 & 0.3022 & 0.7644 & 0.9378 & 0.1733\\
& 2 & 0.3244 & 0.9778 & 0.6533 & 0.4844 & 0.8667 & 0.3822 & 0.5733 & 0.7911 & 0.2178\\
& 4 & 0.2933 & 0.8978 & 0.6044 & 0.3644 & 0.6622 & 0.2978 & 0.4222 & 0.5822 & 0.1600\\
& 8 & 0.2444 & 0.6711 & 0.4267 & 0.2933 & 0.5022 & 0.2089 & 0.3200 & 0.4311 & 0.1111\\
& 16 & 0.2489 & 0.4800 & 0.2311 & 0.2667 & 0.3689 & 0.1022 & 0.2800 & 0.3378 & 0.0578\\
& 32 & 0.2444 & 0.4000 & 0.1556 & 0.2489 & 0.3022 & 0.0533 & 0.2533 & 0.2844 & 0.0311\\ \hline
 \multirow{6}{*}{$20\times 20$}  & 1 & 0.3250 & 1.0000 & 0.6750 & 0.6550 & 0.9825 & 0.3275 & 0.7475 & 0.9400 & 0.1925\\
& 2 & 0.3250 & 0.9825 & 0.6575 & 0.4775 & 0.8475 & 0.3700 & 0.5550 & 0.7600 & 0.2050\\
& 4 & 0.2400 & 0.8925 & 0.6525 & 0.3375 & 0.6650 & 0.3275 & 0.3950 & 0.5750 & 0.1800\\
& 8 & 0.2100 & 0.7025 & 0.4925 & 0.2575 & 0.4625 & 0.2050 & 0.2825 & 0.4025 & 0.1200\\
& 16 & 0.1950 & 0.5325 & 0.3375 & 0.2200 & 0.3325 & 0.1125 & 0.2325 & 0.2900 & 0.0575\\
& 32 & 0.1875 & 0.3150 & 0.1275 & 0.1975 & 0.2625 & 0.0650 & 0.2050 & 0.2400 & 0.0350\\ \hline
 \multirow{6}{*}{$25\times 25$}  & 1 & 0.4400 & 1.0000 & 0.5600 & 0.6832 & 0.9696 & 0.2864 & 0.7824 & 0.9328 & 0.1504\\
& 2 & 0.3040 & 0.9648 & 0.6608 & 0.4672 & 0.8432 & 0.3760 & 0.5552 & 0.7632 & 0.2080\\
& 4 & 0.2448 & 0.8816 & 0.6368 & 0.3216 & 0.6144 & 0.2928 & 0.3744 & 0.5216 & 0.1472\\
& 8 & 0.1776 & 0.7328 & 0.5552 & 0.2336 & 0.4640 & 0.2304 & 0.2640 & 0.3936 & 0.1296\\
& 16 & 0.1600 & 0.4512 & 0.2912 & 0.1888 & 0.3024 & 0.1136 & 0.2048 & 0.2672 & 0.0624\\
& 32 & 0.1536 & 0.3200 & 0.1664 & 0.1680 & 0.2288 & 0.0608 & 0.1760 & 0.2080 & 0.0320\\ \hline 
\multirow{6}{*}{$30\times 30$}  & 1 & 0.4044 & 0.9989 & 0.5944 & 0.6722 & 0.9756 & 0.3033 & 0.7678 & 0.9367 & 0.1689\\
& 2 & 0.2689 & 0.9700 & 0.7011 & 0.4500 & 0.8656 & 0.4156 & 0.5300 & 0.7600 & 0.2300\\
& 4 & 0.2089 & 0.9633 & 0.7544 & 0.2967 & 0.6333 & 0.3367 & 0.3600 & 0.5411 & 0.1811\\
& 8 & 0.1600 & 0.6622 & 0.5022 & 0.2178 & 0.4300 & 0.2122 & 0.2478 & 0.3700 & 0.1222\\
& 16 & 0.1444 & 0.4656 & 0.3211 & 0.1700 & 0.2856 & 0.1156 & 0.1867 & 0.2422 & 0.0556\\
& 32 & 0.1300 & 0.3133 & 0.1833 & 0.1467 & 0.2178 & 0.0711 & 0.1544 & 0.1933 & 0.0389\\ \hline
    \end{tabular}
    \caption{\textbf{Experimental transition values of the probability of getting a 1-DOF $L \times L$ Miura-ori structure under the Most Efficient selection rule}. For each pattern size $L$, we run 500 simulations with an $L \times L$ Miura-ori pattern and calculate the probability $P$ of getting a 1-DOF structure. We then define $\rho^a$ as the maximum $\rho$ with $P = a$ and $\rho_b$ as the minimum $\rho$ with $P = b$ and record their values for different $a,b$ for different number of choices $k$. The difference between every pair of values is also recorded.}
    \label{tab:transition_width}
\end{table*}

\section{Additional analysis of the square case}
\label{sec:appendix_additional}

In the main text, we studied the change in the rigidity percolation transition width in $L\times L$ Miura-ori under the Most Efficient selection rule as the number of choices $k$ increases. In Table~\ref{tab:transition_width}, we present the detailed statistics of the transition width for different pattern sizes $L \times L = 5\times 5$, $10 \times 10$, $15 \times 15$, $20 \times 20$, $25 \times 25$, $30\times 30$ and different number of choices $k=1,2,4,8,16,32$. Specifically, we record the values of $\rho^a$, defined as the maximum $\rho$ with $P = a$, and $\rho_b$, defined as the minimum $\rho$ with $P = b$, and the difference between them.

\begin{figure*}
    \centering
    \includegraphics[width=0.7\linewidth]{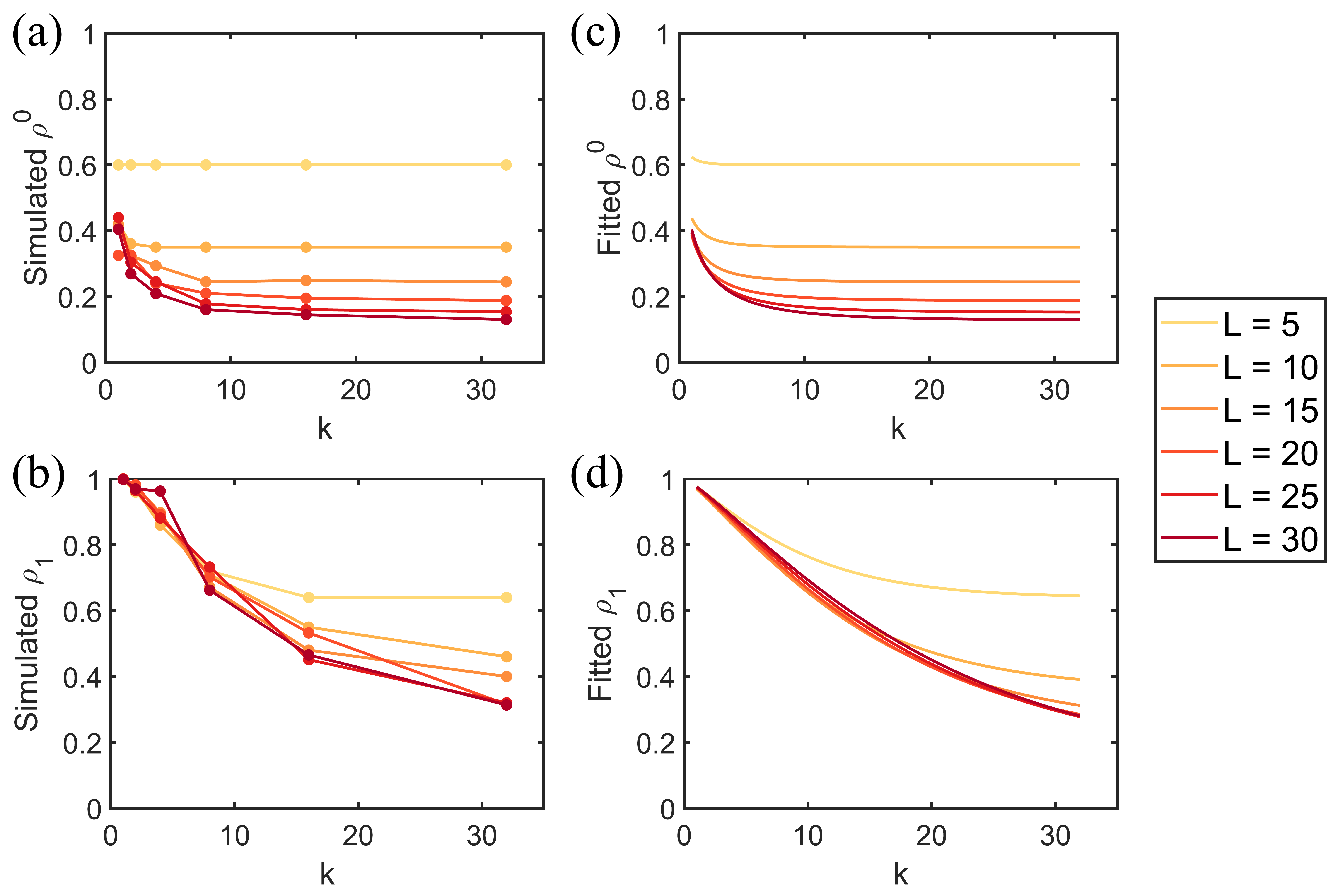}
    \caption{\textbf{The simulated and fitted $\rho^0$ and $\rho_1$ for $L \times L$ Miura-ori structures under the Most Efficient selection rule.} (a)--(b) The simulated $\rho^0$ and $\rho_1$, where each data point represents the result obtained from 500 simulations for a given set of parameters $(L, k)$ (with $L = 5, 10, 15, 20, 25, 30$ and $k = 1,2,4,8, 16, 32$). (c)--(d) The fitted $\rho^0$ and $\rho_1$ obtained using the proposed models in Eq.~\eqref{eqt:fit_rho0} and Eq.~\eqref{eqt:fit_rho1} in the main text.}
    \label{fig:SI_rho0_rho1_sim_vs_fit}
\end{figure*}

It can be observed from the table that both $\rho^0$ and $\rho_1$ decrease generally as $k$ increases. However, their decreasing rates exhibit different behaviors. Specifically, for large pattern size $L \times L$, $\rho^0$ generally decreases faster than $\rho_1$ when the number of choices increases from $k=1$ to some small $k$. This can also be visualized by plotting the simulated values of $\rho^0$ and $\rho_1$ as in Fig.~\ref{fig:SI_rho0_rho1_sim_vs_fit}(a)--(b), from which we can easily see that they show different decreasing rates. Also, as described in the main text, $\rho^0$ and $\rho_1$ can be fitted using two simple models involving a negative exponential term. In Fig.~\ref{fig:SI_rho0_rho1_sim_vs_fit}(c)--(d), we plot the fitted values of $\rho^0$ and $\rho_1$ for different pattern size $L$ and number of choices $k$, from which we can easily see that both formulas match the simulation results very well.

One may ask whether the above observations only hold for the transition interval between $P=0$ and $P = 1$. Besides $\rho^0$ and $\rho_1$, here we also consider the difference between $\rho^{0.1}$ and $\rho_{0.9}$ (i.e. the transition width between $P = 0.1$ and $P = 0.9$) and the difference between $\rho^{0.25}$ and $\rho_{0.75}$ (i.e. the transition width between $P = 0.25$ and $P = 0.75$). As shown in Table~\ref{tab:transition_width}, the differences for these transition intervals also show an increasing trend initially followed by a decreasing trend as $k$ increases. This suggests that our analyses can be naturally extended to other transition intervals.

\section{Additional analysis of general rectangular case}
\label{sec:appendix_mn}

\begin{figure*}[t]
    \centering
    \includegraphics[width=\linewidth]{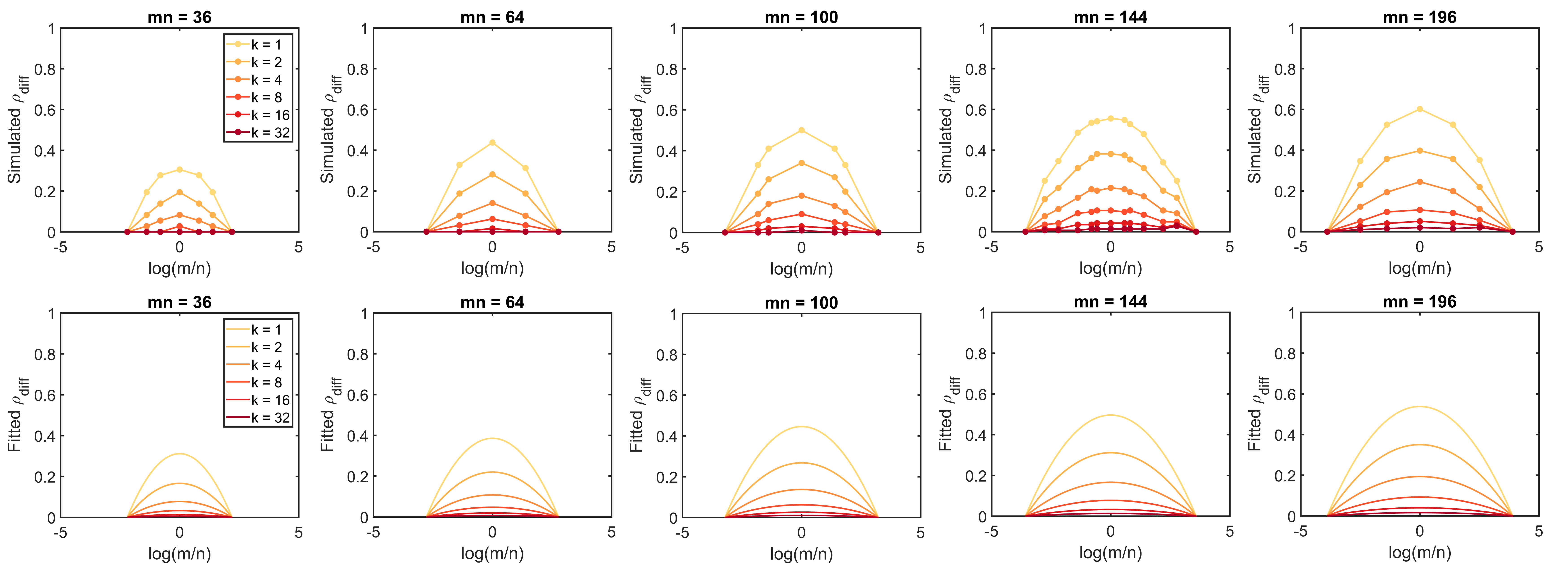}
    \caption{\textbf{The simulated and fitted $\rho_{\text{diff}} = \rho^* - \rho_{\min}$ for general $m \times n$ Miura-ori structures under the Most Efficient selection rule.} The first row shows the simulation results for $mn = 36, 64, 100, 144, 196$, where each data point represents the result obtained from 500 simulations for a specific combination $(m,n,k)$ with $k = 1,2,4,8,16,32$. The second row shows the fitted values of $\rho_{\text{diff}}$ using our proposed model in Eq.~\eqref{eqt:law_diff_rect} in the main text.}
    \label{fig:SI_mn_rdiff}
\end{figure*}

In Eq.~\eqref{eqt:law_diff_rect} in the main text, we gave a simple quadratic model for describing the difference between the critical transition density and the theoretical minimum density $\rho_{\text{diff}} = \rho^* - \rho_{\min}$ for general $m \times n$ Miura-ori structures with different number of choices under the Most Efficient selection rule. In Fig.~\ref{fig:SI_mn_rdiff}, we provide additional examples with $mn = 36, 64, 100, 144, 196$. It can be observed that the simulated $\rho_{\text{diff}}$ is highly symmetric in all examples. Also, we can easily see that the fitted values of $\rho_{\text{diff}}$ using our proposed model match the simulation results very well.

\begin{figure*}[t]
    \centering
    \includegraphics[width=\linewidth]{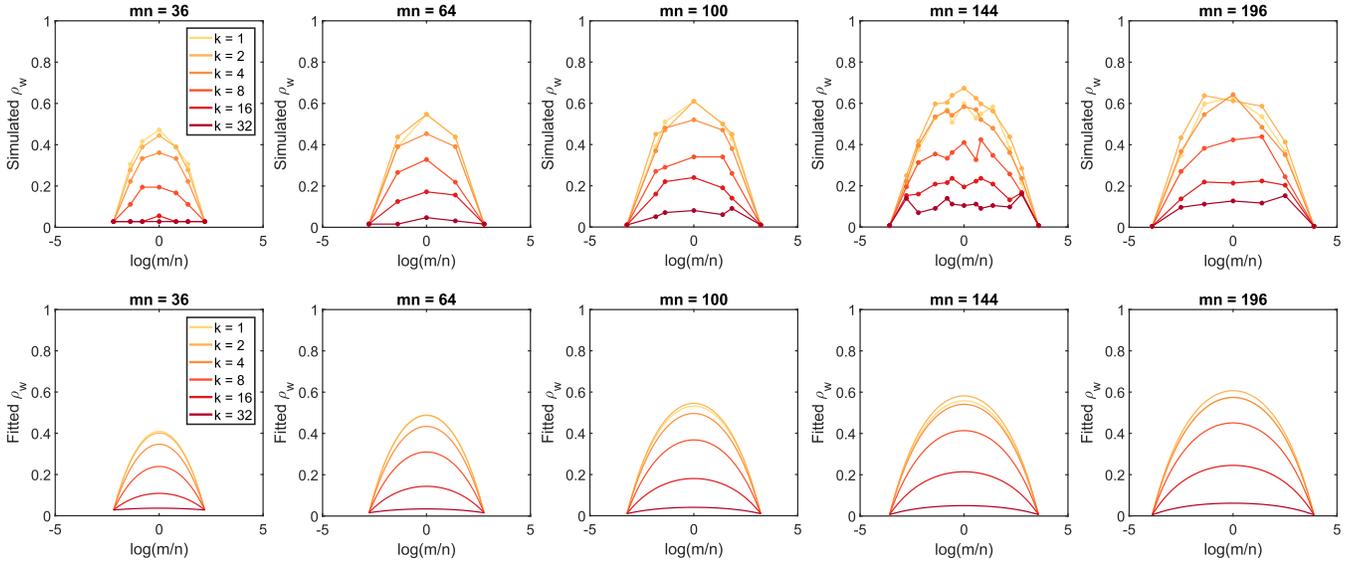}
    \caption{\textbf{The simulated and fitted rigidity percolation transition width $\rho_w$ for general $m \times n$ Miura-ori structures under the Most Efficient selection rule.} The first row shows the simulation results for $mn = 36, 64, 100, 144, 196$, where each data point represents the result obtained from 500 simulations for a specific combination $(m,n,k)$ with $k = 1,2,4,8,16,32$. The second row shows the fitted values of $\rho_w$ using our proposed model.}
    \label{fig:SI_mn_rwidth}
\end{figure*}

Also, we proposed a simple approximation formula for the rigidity percolation transition width $\rho_w$ for any given $(m,n,k)$ for general $m\times n$ Miura-ori structures in the main text. In Fig.~\ref{fig:SI_mn_rwidth}, we provide additional examples with $mn = 36, 64, 100, 144, 196$ for visualizing the change of $\rho_w$ with different combinations of $(m,n,k)$. Again, we can see that the fitted $\rho_w$ using our proposed model largely resembles the simulated transition width for all $mn$.

\begin{figure}[t!]
    \centering
    \includegraphics[width=0.8\linewidth]{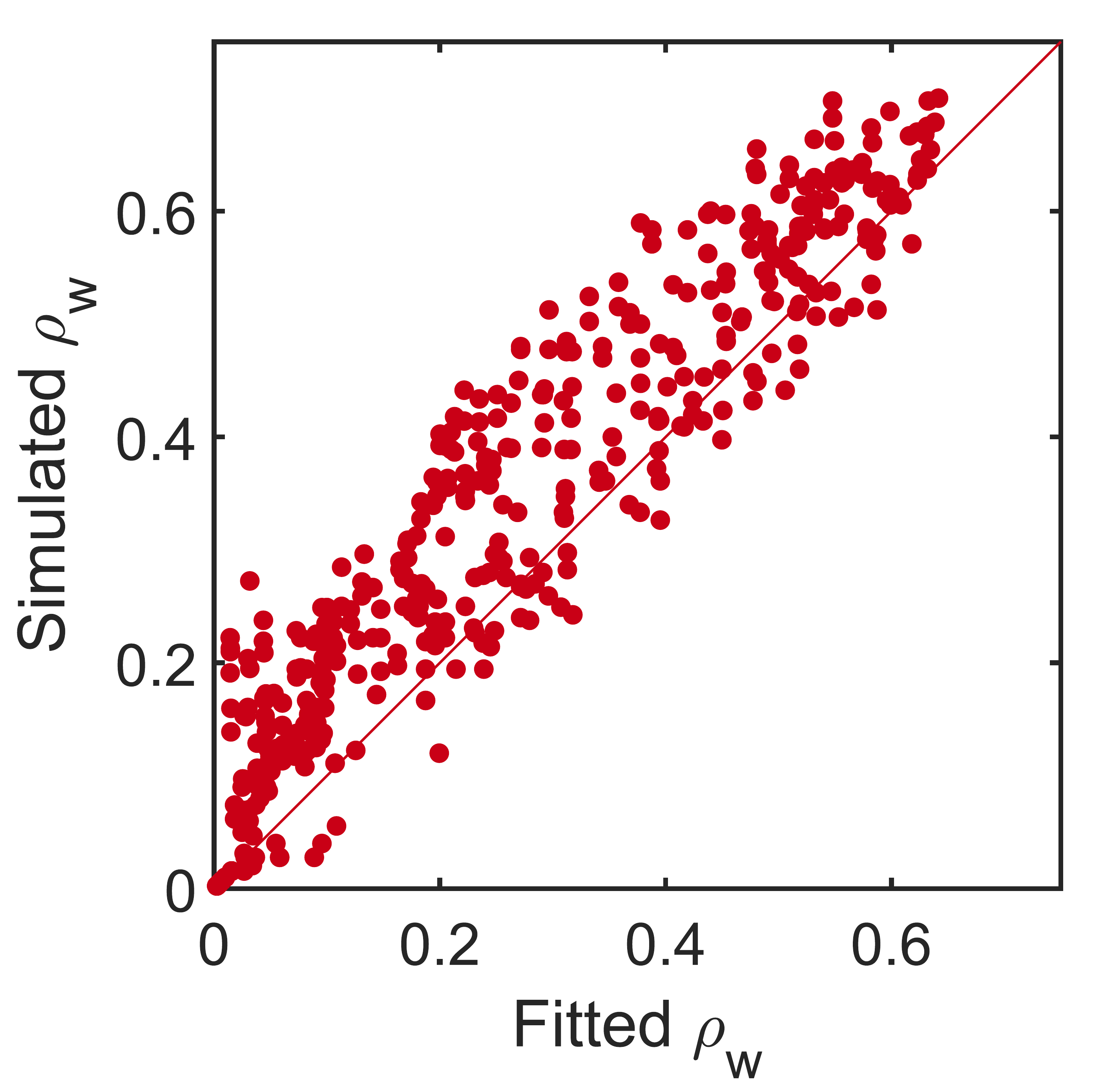}
    \caption{\textbf{A comparison between the simulated rigidity percolation transition width $\rho_w$ and the alternative model in Eq.~\eqref{eqt:law_width_rect_2} for $m\times n$ Miura-ori structures.} Each data point represents the fitted value ($x$-coordinate) and simulated value ($y$-coordinate) for a combination of $(m,n,k)$, with $mn = 25, 36, 49, 64, \dots, 400$ and $k = 1,2,4,8,16,32$. The red line represents $y = x$.}
    \label{fig:SI_rect_sim_vs_fit2}
\end{figure}

Besides, as $\rho_w$ is roughly symmetric about $\log(m/n) = 0$, it is natural to ask whether one can also approximate $\rho_w$ using a simple quadratic polynomial in $\log(m/n)$ analogous to the one for $\rho_{\text{diff}}$ in the main text. Here, we consider the following alternative approximation formula for $\rho_w$:
\begin{equation} \label{eqt:law_width_rect_2}
\begin{split}
    & \rho_w(m,n,k) \\\approx &\left({\rho}_w(L,k) - \frac{1}{mn}\right) \left(1 - \left(\frac{\log (m/n)}{\log (mn/4)}\right)^2\right) + \frac{1}{mn},
\end{split}
\end{equation}
where ${\rho}_w(L,k)$ is the fitted value for the square case $L \times L$ with $L = \sqrt{mn}$ using main text Eq.~\eqref{eqt:law_width}. Specifically, note that for $m=n$, we have $\log (m/n) = 0$ and hence 
\begin{equation}
    \left({\rho}_w(L,k) - \frac{1}{mn}\right) (1 -0)^2 + \frac{1}{mn} = {\rho}_w(L,k).
\end{equation}
In other words, Eq.~\eqref{eqt:law_width_rect_2} is identical to Eq.~\eqref{eqt:law_width} in the main text if $m=n$. Also, it is easy to see that Eq.~\eqref{eqt:law_width_rect_2} is perfectly symmetric and gives the same value for $(m,n,k)$ and $(n,m,k)$, which matches our observation. Moreover, for the extreme case where $m = 2$ or $n=2$, note that all facets of the Miura-ori structure are boundary facets and we need to explicitly add the quad planarity constraints to all of them to make the structure 1-DOF. This requires exactly $mn$ steps and hence the theoretical transition width will be $1/(mn)$. Now, if $n = 2$, we have
\begin{equation}
\begin{split}
    1 - \left(\frac{\log (m/n)}{\log (mn/4)}\right)^2 &= 1 - \left(\frac{\log (mn/2)/2)}{\log (mn/4)}\right)^2 \\
    &= 1 - 1 = 0,
\end{split}
\end{equation}
and it follows from Eq.~\eqref{eqt:law_width_rect_2} that
\begin{equation}
    \rho_w \approx 0 + \frac{1}{mn} = \frac{1}{mn}.
\end{equation}
Similarly, if $m = 2$, we have $\rho_w \approx 1/(mn)$. This shows that Eq.~\eqref{eqt:law_width_rect_2} matches the expected $\rho_w$ at the peak and the two endpoints. In Fig.~\ref{fig:SI_rect_sim_vs_fit2}, we compare the fitted values of $\rho_w$ by this simple quadratic model and the simulation results for different combinations of $(m,n,k)$ as in the main text. It can be observed that this alternative model gives a qualitatively good fit, with most data points being close to the line $y = x$. However, comparing this plot with the plot for the original model in main text Fig.~\ref{fig:analysis_rect}(c), we can see that the original model in main text Eq.~\eqref{eqt:law_width_rect} gives a better fit with a smaller deviation from $y=x$. 

Altogether, this alternative model provides a simple and reasonably accurate way to connect the rigidity percolation transition width of $L \times L$ Miura-ori structures with that of all other Miura-ori structures with the same total number of facets.

\end{document}